\documentclass{aa}
\usepackage{graphicx}
\usepackage{subcaption}
\usepackage{enumitem}
\usepackage[dvipsnames]{xcolor}
\usepackage{amsmath}
\usepackage{txfonts}
\usepackage{soul}
\usepackage[colorlinks=True,pdfborder={0 0 0}, linkcolor=blue,citecolor=blue]{hyperref}
\begin{document}

   \title{Birth of magnetized low-mass protostars and circumstellar disks}

   \author{A. Ahmad
          \inst{1,3}
          \and
          M. González\inst{1}
          \and
          P. Hennebelle\inst{2}
          \and
          U. Lebreuilly\inst{2}
          \and
          B. Commerçon\inst{3}
          }

   \institute{Université Paris Cité, Université Paris-Saclay, CEA, CNRS, AIM, F-91191, Gif-sur-Yvette, France
             \and
             Université Paris-Saclay, Université Paris Cité, CEA, CNRS, AIM, 91191, Gif-sur-Yvette, France
             \and
             Univ Lyon, Ens de Lyon, Univ Lyon 1, CNRS, Centre de Recherche Astrophysique de Lyon UMR5574, 69007, Lyon, France
             }

   \date{Received XXXX; accepted XXXX}

 
  \abstract
  {Providing a comprehensive description of the birth of protostars and circumstellar disks, and how these two evolve over time, is among the goals of stellar formation theory. Although the two objects are often studied separately owing to numerical and observational challenges, breakthroughs in recent years have highlighted the need to study both objects in concert. The role of magnetic fields in this regard must also be investigated, and current observational surveys broadly report $\sim \mathrm{kG}$ field strengths in young stellar objects.}
   {We aim to describe the birth of the protostar and that of its circumstellar disk, as well as their early joint evolution following the second collapse. We wish to study the structure of the nascent star-disk system, and that of its magnetic fields, while focusing on the innermost sub-AU region.}
   {We carry out very high resolution 3D radiative-magnetohydrodynamics simulations, describing the collapse of turbulent dense cloud cores to stellar densities, both under the ideal and non-ideal approximation in which ambipolar diffusion is accounted for. The calculations are integrated as far as possible in time, reaching $\approx 2.3$ yr after protostellar birth. Our simulations are also compared to their hydrodynamical counterpart to better isolate the role of magnetic fields.}
   {In line with previous results, we find that the ideal MHD run yields extremely efficient magnetic braking, which suppresses the formation of circumstellar disks and produces a central, spherical protostar. In addition, this run predicts a magnetic field strength of $\sim 10^{5}$ G within the protostar at birth. In the non-ideal run, the efficiency of magnetic braking is drastically reduced by ambipolar diffusion and the nascent protostar reaches breakup velocity, thus forming a rotationally supported circumstellar disk. The diffusion of the magnetic field also allows for the implantation of a $\sim \mathrm{kG}$ field in the protostar, which is thereafter maintained. The magnetic field is mainly toroidal in the star-disk system, although a notable vertical component threads it. No outflows or jets are reported owing to our use of turbulent initial conditions, which reduces the coherence of the magnetic field, although we report that conditions are being set in place for it to occur at later times. We also show that the nascent circumstellar disk is prone to the magneto-rotational instability, although our resolution is inadequate to capture the mechanism. We note a sensitivity of the nascent disk's properties with regards to the angular momentum inherited prior to the dissociation of H$_2$ molecules, as well as the magnetic field strength, thus stressing the need for better constraints on dust resistivities throughout the collapse.}
   {These calculations illustrate the role of magnetic fields in dictating the behavior of the gas throughout the collapse. They carry multiple implications on several issues in stellar formation theory, and offer perspectives for future modeling of the innermost regions of the star-disk system. Most notably, should the fossil field hypothesis to explain the origins of magnetic fields in young stellar objects hold, we show that a $\sim \mathrm{kG}$ field strength may be implanted and maintained in the protostar at birth.}


   \maketitle
%

\section{Introduction}
   In recent years, the role of magnetic fields in star formation has garnered a significant amount of interest. Aided by advances in far-infrared and submillimeter (e.g., ALMA, NOEMA, VLA) instruments capable of measuring linearly polarized dust emissions, magnetic fields have been observed in dense cloud cores \citep{kirk_2006, jones_2015, kandori_2018, meyers_2021} where they exhibit supercriticality (i.e., the mass-to-flux ratio is above unity) and a typical field strength of $\sim 10^{-5}$ G. Furthermore, using Zeeman line splitting techniques \citep{crutcher_2019}, they have also been observed in young stellar objects with values of $\sim 10^{3}$ G \citep{johns_2007, johns_2009, yang_2011, flores_2024}. Should the magnetic field be perfectly coupled to the fluid during the collapse of the dense core (i.e., the ideal MHD approximation), flux freezing implies that the resulting protostar would have a magnetic field strength of $\sim 10^{6}$ G, far in excess of observed values. Therefore, a considerable amount of magnetic flux is lost by the time the protostar becomes visible. This problem is known as the \textit{magnetic flux problem} \citep{mouschovias_1985}, which has so far eluded a concise answer. Current observational surveys of magnetic fields of Young Stellar Objects (YSOs), although limited in sample size, have so far failed to find any correlation between magnetic field strength and stellar properties, such as their age and rotational period. However, they report a decreasing magnetic flux over time \citep{yang_2011}. 
   \\
   The origin of the observed magnetic fields in YSOs is currently a subject of debate. Two main hypothesis dominate the discourse; the fossil field hypothesis, whereby the measured magnetic fields in these evolved sources are carried over from their inception in the second Larson core\footnote{The object in hydrostatic equilibrium formed after the collapse caused by the dissociation of H$_2$ molecules (i.e., the nascent protostar, \citealp{larson1969})}, and the dynamo hypothesis, which states that the measured fields are produced through a dynamo process. Ultimately, solving this problem requires a detailed model of the evolution of the protostellar magnetic field as the protostar transitions from the Class 0 to the Class I phase, accounting for prestellar evolution and describing the magnetic field's evolution using dynamo theory, however such a model is yet to be developed and little is reported on the subject in the literature. In the absence of any such model, the fossil field hypothesis remains favored and this may provide a constraint on star formation simulations, as they must be able to form a protostar whose magnetic field has a strength of $\sim 10^{3}$ G.
\\
\\
A similar and closely linked issue to this is the \textit{angular momentum problem}, which states that should angular momentum be conserved during the collapse of the dense core, stars would rotate far above their breakup velocity and circumstellar disks would be an order of magnitude larger than their observed sizes ($\sim 30^{1}$ AU, \citealp{maury_2019, tobin_2020}). Once again, the ideal MHD approximation fails to conform to observational data as it produces magnetic braking that is efficient enough to extract all angular momentum from the dense core (i.e., the \textit{magnetic braking catastrophe}, \citealp{matsumoto_2004, hennebelle_2008, hennebelle_2008b, mellon_2008}). It is now widely admitted that resistive processes, mainly ambipolar diffusion, are responsible for breaking the ideal MHD limit towards higher density gas and reducing the magnetic braking efficiency to the point where a disk may form and reach sizes comparable to observations (e.g., \citealp{masson_2016, hennebelle_2016, vaytet_2018, machida_2019, wurster_disks, wurster_2020, mayer_2024}, see additionally the review by \citealp{tsukamoto_2023}). In addition to this, resistive MHD simulations report a converged magnetic field strength of $\sim 0.1$ G in the first Larson core (primarily due to ambipolar diffusion), which allows for the second Larson core to form with a magnetic field strength of $\sim 10^3$ G (e.g., \citealp{vaytet_2018, machida_2019, wurster_2022, mayer_2024}).
\\
In order to account for resistive processes, one must use a detailed chemical network that describes the abundance of charged species. In addition, one must also make an assumption on the dust grain size and density distribution in order to determine the surface area available for chemical reactions \citep{marchand_2016, zhao_2020, marchand_2021, marchand_2022}, and to account for the fact that the dust particles themselves may be the main charge carriers. In this regard, the Mathis-Rumple-Nordsiek distribution (MRN, \citealp{mathis_1977}), or some of its variants, is most often used as it is mostly valid for dust particles in the interstellar medium. However, recent studies having undertaken the effort of re-evaluating the dust size distribution during the collapse of dense cores have called into question the validity of the MRN distribution \citep{guillet_2020, silsbee_2020, lebreuilly_2023c, kawasaki_2023, tsukamoto_2023, goy_2024}. Most notably, these studies reveal an absence of small grains toward gas densities close to first Larson core values ($\sim 10^{-13}\ \mathrm{g\ cm^{-3}}$), which in turn causes a stark drop in Ohmic resistivity, to the point where it is no longer a viable dissipative process at densities of the first Larson core and higher. The Hall effect, even within the MRN framework, remains the most poorly constrained resistive effect. Studies accounting for it report drastically different evolutionary scenarios for protoplanetary disks (e.g., \citealp{tsukamoto_2015b, wurster_disks, wurster_2022}), however the computed resistivities are too uncertain to draw any conclusions on the subject. The only resistive effect whose role and behavior during the protostellar collapse can be inferred with some confidence is ambipolar diffusion, whose reduction in magnetic braking efficiency allows for the formation of disks whose sizes are in broad agreement with Class 0 disk size surveys \citep{hennebelle_2016, maury_2019, tobin_2020}. However, it still varies by orders of magnitude under different assumptions of dust coagulation.
\\
\\
Finally, it has recently become clear that subgrid models wrapped into sink particles that are used in protostellar collapse calculations in order to alleviate timestepping constraints produce results that are very sensitive to their parameters \citep{machida_2014, vorobyov_2019, hennebelle_disks}. As said sink particles have a wide field of applications (e.g., \citealp{krumholz_2009, kuiper_2010, bate_2012, bate_2018, hennebelle_2020, mignon_2021b, mignon_2021a, lebreuilly_2021, commercon_2022, grudic_2022, lebreuilly_2023, lebreuilly_2023b, mignon_2023, andre_2023, kuruwita_2024}), it is of vital interest to adequately constrain the subgrid parameters used in them. This requires one to study the innermost sub-au region of circumstellar disks, which entails second-collapse calculations resolving both the protostar and the newly-formed circumstellar disk. 
\\
\\
These second-collapse calculations can link the aforementioned issues together by providing a direct prediction of the magnetic flux and angular momentum inherited by the protostar, as well as by describing the star-disk interaction in detail.
A number of said calculations exist in the literature (e.g., \citealp{banerjee_2006, tomida_2013, vaytet_2018, machida_2019, wurster_2020, wurster_2022, mayer_2024}). However, the majority of these use idealized setups in which solid-body rotation is assumed and turbulence is absent in the initial dense cloud core. This absence of turbulence allows the magnetic field to maintain a coherence which amplifies its effects, be it magnetic braking or the launching of outflows and jets. In addition, with the exception of \cite{machida_2019, wurster_2020, wurster_2022} (which use nested-grid or SPH methods), these calculations are often stopped soon after protostellar birth owing to timestepping constraints, and the studies that have undertaken the challenge of integrating further out in time have not provided an in-depth analysis of the gas kinematics in the innermost sub-AU region. Nevertheless, they found that the resulting protostar and circumstellar disk are thermally supported bodies, where thermal pressure gradient forces vastly outweigh their magnetic counterparts. In this regard, \cite{ahmad_2024} let a study in which the RHD approximation was used and found that the nascent protostar quickly reaches breakup speeds, by which point a circumstellar disk forms around it and expands outward. This occurs regardless of the initial conditions of the parent dense core. \cite{machida_2011b} and \cite{bhandare_2024} similarly report the existence of this disk surrounding the protostar (hereafter called the inner disk or circumstellar disk).
\\
Recent observational studies have also just begun probing deeply embedded Class 0 sources \citep{laos_2021, gouellec_2024, gouellec_2024b}, and although the structure of the system at the innermost sub-AU region is as of yet difficult to infer from said observations, these seem to be reporting vigorous accretion onto the newly formed protostar.
\\
\\
In the present study, we expand upon our work in \cite{ahmad_2024} by including the effects of magnetic fields in our calculations, both under the ideal and non-ideal MHD approximation, while accounting for radiative transfer using the the Flux Limited Diffusion (FLD) approximation and including turbulence in our initial dense cloud core. Our goals are to describe the birth and early evolution of the protostar and the circumstellar disk surrounding it by focusing on the innermost sub-AU region. The simulations presented in this paper have the highest effective 3D resolution of all second-collapse RMHD simulations, all-the-while pushing the calculations as far as possible in time. In light of the aforementioned recent studies around dust size distribution during protostellar collapses that report a stark drop in Ohmic resistivities, we have chosen to ignore Ohmic dissipation in our non-ideal MHD simulation, and only ambipolar diffusion is accounted for. We follow the collapse of a dense cloud core to stellar densities, and describe the initially isothermal phase of the collapse, the formation of the first Larson core and its subsequent adiabatic contraction, the second collapse following the dissociation of H$_2$ molecules, the birth of the protostar, and subsequently push the calculations as far as possible in time. In our pursuit of describing the smallest spatial scales relevant to protostellar and circumstellar disk birth, we have obtained the best resolved protostars and circumstellar disk in the MHD literature. Particular attention is given to the structure of the magnetic field within the nascent protostar, as well as within the circumstellar disk. The evolution of the nascent circumstellar disk is also compared to its hydro counterpart in order to better ascertain the effects of magnetic fields on the system.
\\
Our results, reported below, carry multiple implications for the angular momentum and the magnetic flux problems. In addition, they offer constraints on subgrid parameters used in disk evolution studies.
\\
In \hyperref[section:model]{Sec. \ref*{section:model}}, we present our numerical and physical setup, whose results are discussed in \hyperref[section:results]{Sec. \ref*{section:results}}. The implications of our results are discussed in \hyperref[section:discussions]{Sec. \ref*{section:discussions}}. Finally, we conclude in \hyperref[section:conclusion]{Sec. \ref*{section:conclusion}}

\section{Model} \label{section:model}
\subsection{Numerical setup}

We have used the {\ttfamily RAMSES} astrophysical code \citep{teyssier_2002} and its extension to MHD by \cite{Fromang_2006}. Non-ideal MHD (ambipolar + Ohmic dissipation) was implemented in the code by \cite{masson_2012}, and radiative transfer under the flux limited diffusion approximation was implemented by \cite{commercon_2011, commercon_2014, gonzalez_2015}. The equation of state and opacity tables used were pieced together by \cite{vaytet_2013} from \cite{saumon_1995, semenov_2003, ferguson_2005}, and  \cite{badnell_2005}. These account for a gas mixture consisting of 73\% hydrogen and 27\% helium, and assume a 1\% dust mass content within the gas.
\\
We use the same initial setup as in \cite{ahmad_2024}, with the same refinement strategy. It consists of a uniform density dense cloud core of radius $R_{0}=2.465\times 10^{3}$ AU, mass $M_{0}=1\ \mathrm{M_{\odot}}$ and temperature $10$~K, equivalent to an initial ratio of thermal to gravitational energy of 0.25. Angular momentum is present through the inclusion of a turbulent velocity vector field parameterized by the turbulent Mach number $\mathcal{M}$, which we have set to 0.4. Radiative transfer is accounted for under the gray FLD approximation. A uniform magnetic field threads the dense cloud core along the $z$ axis, and its strength is parameterized by the mass-to-flux ratio which we have set to 4. This corresponds to an initial magnetic field strength of $\sim 10^{-5}$~G in the dense cloud core, and an Alfvénic Mach number of $\mathcal{M}_{\mathrm{a}}\approx 0.12$. This setup is identical to that of run G2 in \cite{ahmad_2024}, with the only difference being the presence of magnetic fields.
\\
Two simulations will be presented in this section; one under the ideal MHD approximation (hereafter IMHD) and one in which we have accounted for ambipolar diffusion (hereafter NIMHD). Run IMHD is mainly presented in this paper for comparative purposes with the more realistic run NIMHD, as that allows us to better isolate the role of magnetic resistivities during the simulation.
\\
These two simulations use the same refinement strategy, however run IMHD has a maximum refinement level of $\ell_{\mathrm{max}}=26$ (the coarsest level is at $\ell_{\mathrm{min}}=6$) whereas run NIMHD has $\ell_{\mathrm{max}}=25$. As we will see later-on, this is because run IMHD forms a much more compact protostar, whose properties require a finer spatial resolution to describe. This means that at the finest refinement level, run IMHD and NIMHD respectively have a spatial resolution of $\Delta x_{\mathrm{IMHD}} = 1.4\times 10^{-4}$~AU and $\Delta x_{\mathrm{NIMHD}} = 2.9\times 10^{-4}$~AU.

\subsubsection{Zoom-out}
\label{sec:zoomout}

Owing to very stringent time-stepping constraints following protostellar birth, run NIMHD requires approximately two days of CPU wall time in order to integrate $\approx 40$ hours. This is because the timestep at the finest level reaches a mere minute, and the poor load balancing causes most cells to be handled by a few CPUs. As the protostar and circumstellar disk grow and expand over time, this problem is aggravated as a considerable number of cells are created to describe the newly formed structures. In order to alleviate the timestepping constraints, we have also run a simulation branched out of run NIMHD nearly 0.4 years after protostellar birth, in which the maximum refinement level was reduced from $\ell_{\mathrm{max}} = 25$ to $\ell_{\mathrm{max}}=24$. This allowed us to push the simulation considerably further out in time,\footnote{The $\ell_{\mathrm{max}}=25$ simulation ran for $\approx 0.55$ years after protostellar birth, whereas the $\ell_{\mathrm{max}}=24$ simulation ran from $\approx 0.44$ years to $\approx 2.33$ years.} which is particularly useful to study the expansion of the newly-formed circumstellar disk. This run, labeled "NIMHD\_LR", is discussed in \hyperref[sec:diskexpansion]{Sec. \ref*{sec:diskexpansion}}.
\\
\\
Run IMHD required 2 months and 4 days of CPU wall time, whereas run NIMHD required 7 months and 6 days. Run NIMHD\_LR ran for 6 months and 13 days from its branch-out point, meaning that when including the zoom-out, the non-ideal simulation ran for over a year of CPU wall time. All simulations were run on 64 CPU cores.

\section{Results} \label{section:results}

\subsection{Large scale structures}

\begin{figure*}[h]
\centering
\includegraphics[scale=.5]{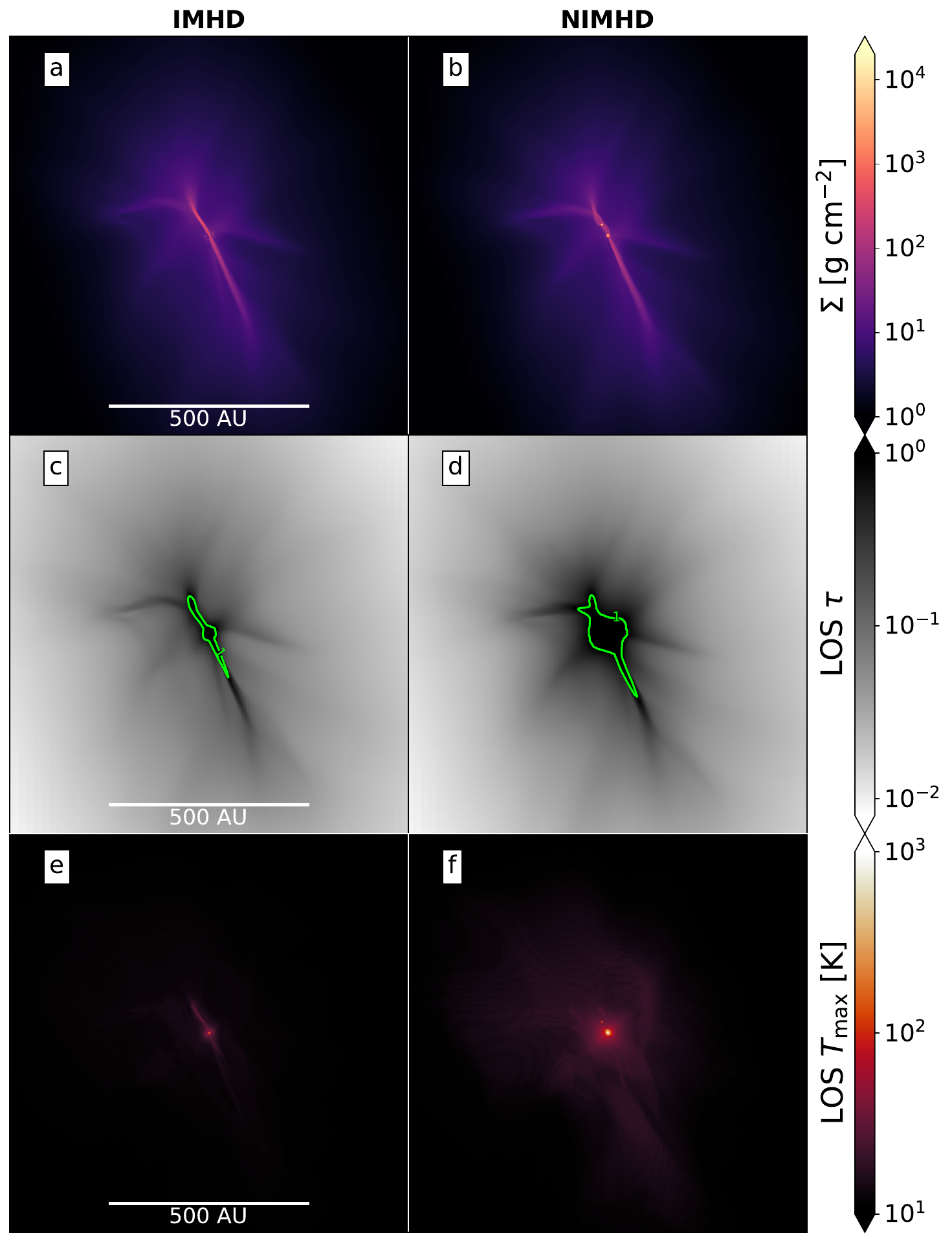} 
\caption[Large scale structure comparison between runs IMHD and NIMHD.]{A comparison of runs IMHD (first column) and NIMHD (second column) at the scale of the dense cloud core itself. The snapshots are taken respecitvely at $t\approx 23.25$ and $t\approx 23.39$ kyr following the collapse of the dense core. The first row displays column density (panels a and b), the second row displays the optical depth computed along the line of sight (panels c and d), and the last row displays the maximum temperature along the line of sight (panels e and f). All maps are projections along the $z$ axis. The lime-colored contour in panels (c) and (d) represent an optical depth of unity. The scale bars in the first column apply to the second column as well.}
\label{fig:largescale}
\end{figure*}

\begin{figure*}[h]
\centering
\includegraphics[scale=.35]{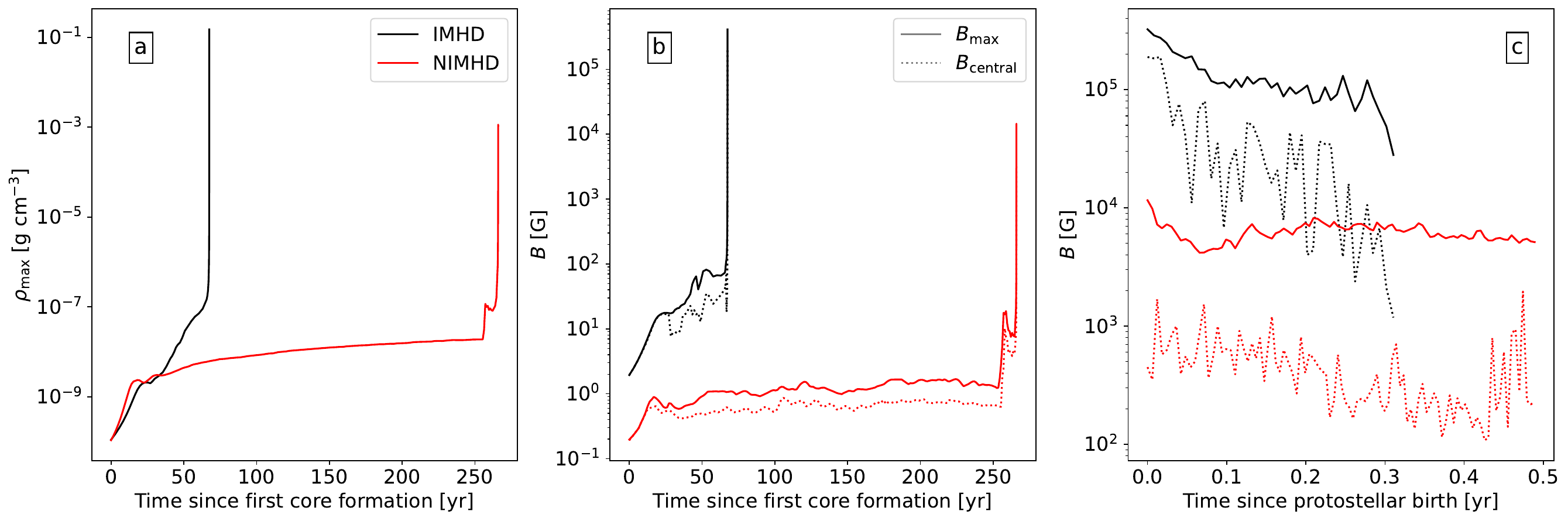} 
\caption[Magnetic field properties during the collapse]{A quantitative comparison of the collapse between run IMHD (black) and run NIMHD (red). Panel (a) displays the evolution of the maximum density since the formation of the first Larson core, which we define as the time where a density of $\approx 10^{-10}\ \mathrm{g\ cm^{-3}}$ is achieved. Panel (b) and (c) display the magnetic field strength evolution as a function of time since first core formation and since protostellar birth (defined as the moment a density of $\approx 10^{-5}\ \mathrm{g\ cm^{-3}}$ is reached), where the solid lines represent the maximum magnetic field strength and the dotted lines represent the field's strength measured at the location of maximum density.}
\label{fig:coresinfo}
\end{figure*}

We first begin by describing the system at the scale of the dense core itself ($\sim 10^{3}$ AU), with the goal of providing the contextual environment in which the protostar is born. To this end, we compare runs IMHD and NIMHD in \hyperref[fig:largescale]{Fig.~\ref*{fig:largescale}} at our final simulation snapshots (respectively $\approx 23.25$ and $\approx 23.39$ kyr after simulation start), which displays the column density (first row), the optical depth $\tau$ computed along the line of sight (second row), and the maximum temperature along the line of sight (third row). $\tau$ is computed as
\begin{equation}
    \tau = \int_{z_{\mathrm{min}}}^{z_{\mathrm{max}}} \rho \kappa_{\mathrm{R}} dz\ ,
\end{equation}
where $\rho$ is the gas density and $\kappa_{\mathrm{R}}$ the Rosseland mean opacity.
\\
The column density maps show a filamentary structure of size $\sim 10^{2}$ AU forming in both runs. This structure is formed by gravo-turbulence \citep{tsukamoto_2013}, however it appears much thinner in these calculations than their RHD counterparts in \cite{ahmad_2024}. This is due to magnetic braking, which extracts a significant amount of angular momentum from the gas and thus prevents it from spreading out as much. Since ambipolar diffusion begins acting at higher densities ($\sim 10^{-14}\ \mathrm{g\ cm^{-3}}$), the two runs yield identical column density maps outside the filament, however in the case of run NIMHD it has fragmented into two distinct dense cores \citep{fiedge_2000}. The existence of a secondary bound fragment within the filament is owing to an extended first core lifetime. Indeed, the first core survived $\approx 100$ years longer in run NIMHD owing to a reduced mass accretion rate onto it, which is in turn due to less efficient magnetic braking. In this time span, the filament fragmented in run NIMHD, whereas the stringent timestepping following the second collapse froze the simulation at larger scales in run IMHD, and no bound fragment is witnessed at its final simulation snapshot.
\\
Despite the very similar structure, the two runs have differing optical depth maps (\hyperref[fig:largescale]{Fig.~\ref*{fig:largescale}} c and d). Indeed, run NIMHD has a more spatially extended optically thick region (lime-colored contours) than run IMHD. Since the two runs display identical column density values at the location where said optical thickness is achieved in run NIMHD, the differing optical depth maps are due to the differing temperatures found within the cloud core, as ambipolar diffusion significantly heats-up the gas (\hyperref[fig:largescale]{Fig.~\ref*{fig:largescale}} f). The increase in temperature at these densities manifests itself as an increase in opacity (see figure 1 of \citealp{ahmad_2023}). This serves to show that the two models should produce distinct emission maps that may be discriminated against with current observational instruments.
\\
\\
We now turn to studying the collapse in quantitative terms using \hyperref[fig:coresinfo]{Fig.~\ref*{fig:coresinfo}}. \hyperref[fig:coresinfo]{Figure~\ref*{fig:coresinfo}} (a) displays the maximum density of the simulation as a function of time since first core formation (defined as the moment where $\rho_{\mathrm{max}}>10^{-10}\ \mathrm{g\ cm^{-3}}$). The steep rise in $\rho_{\mathrm{max}}$ in this figure corresponds to the second collapse (i.e., protostellar birth). We see here that the two runs display different first core lifetimes, with run NIMHD entering the second collapse phase nearly 200 years later. This is in contrast to \cite{vaytet_2018}'s results, who reported a longer first core lifetime in their ideal MHD simulation due to the interchange instability reducing mass accretion rates onto it. This discrepancy between our results is likely due to our use of turbulent initial conditions, which although does not prevent the emergence of the interchange instability in run IMHD (\hyperref[fig:largescale]{Fig.~\ref*{fig:largescale}} a), still reduces its efficiency.\footnote{\cite{vaytet_2018} also had slightly more stable initial conditions, as their thermal-to-gravitational energy ratio is 0.28, whereas ours is 0.25.} The first core in run NIMHD survived for a total of $\approx 250$ years, which is about half as much as the hydrodynamical run presented in \cite{ahmad_2024}. Its extended lifetime in comparison to run IMHD is due to the reduced magnetic braking efficiency, which allows for angular momentum to reduce mass accretion rates onto the first core. The maximum density reached post-second collapse in run IMHD is $\sim 10^{-1}\ \mathrm{g\ cm^{-3}}$, and $\sim 10^{-3}\ \mathrm{g\ cm^{-3}}$ in run NIMHD.
\\
During the collapse, flux freezing causes the magnetic field strength to increase with increasing density (with $B\propto \rho^{2/3}$). This causes the maximum magnetic field strength ($B_{\mathrm{max}}$) in run IMHD, shown in \hyperref[fig:coresinfo]{Fig.~\ref*{fig:coresinfo}} (b), to continuously increase over time (with increasing central density), with small drops in magnetic field strength being caused by turbulent reconnection. In run NIMHD however, the magnetic field strength displays a plateau at $\approx 0.5$ G, owing to ambipolar diffusion. This field strength within the first core has been consistently retrieved by numerous studies in the literature that account for ambipolar diffusion, both in the low-mass and in the high-mass regime (e.g., \citealp{masson_2016, vaytet_2018, mignon_2021a, wurster_2022, mayer_2024}). Once the second collapse occurs, flux freezing (which is recovered in run NIMHD following dust sublimation and the ionization of atomic gas species) once again causes a strong increase in magnetic field strength, which reaches $\sim 10^{5}$~G in run IMHD and $\sim 10^{3}$~G in run NIMHD. We also notice in both runs that the magnetic field strength measured at the location of maximum density ($B_{\mathrm{central}}$, dotted lines) is a factor $\approx 2$ below $B_{\mathrm{max}}$. Following the second gravitational collapse (\hyperref[fig:coresinfo]{Fig.~\ref*{fig:coresinfo}} c), the maximum magnetic field strength reaches $\approx 3\times 10^{5}$~G in run IMHD, and $\approx 10^{4}$~G in run NIMHD. Soon after protostellar birth, the maximum field strength in run IMHD continuously decreases to $\sim 5\times 10^{4}$~G, and in the case of run NIMHD, it decreases to $\sim 6\times 10^{3}$~G and plateaus around this value. We see the same trend in $B_{\mathrm{central}}$, which fails to coincide with $B_{\mathrm{max}}$ following the second collapse, and whose discrepancy with it seems to worsen over time. We show later in \hyperref[sec:magfieldstruct]{Sec. \ref*{sec:magfieldstruct}} that the drop in magnetic field strength in run IMHD is mostly due to an outward advection of magnetic flux. The discrepancy between $B_{\mathrm{max}}$ and $B_{\mathrm{central}}$ has been reported in previous papers in the literature, most notably \cite{wurster_2020, wurster_2022}. They also report a reduction in $B_{\mathrm{max}}$ shortly following protostellar birth. Our results confirm their findings, however the cells containing $\rho_{\mathrm{max}}$ and $B_{\mathrm{max}}$ are separated by a very small distance ($\sim 10^{-2}$~AU) in our simulations, whereas in theirs it is of the order of $\sim 1$ AU, and we do not report the existence of a "magnetic wall" on which magnetic flux is accumulated as they do.

\subsection{The second collapse}

\subsubsection{Qualitative result of the second collapse}

\begin{figure*}[h]
\centering
\includegraphics[scale=.28]{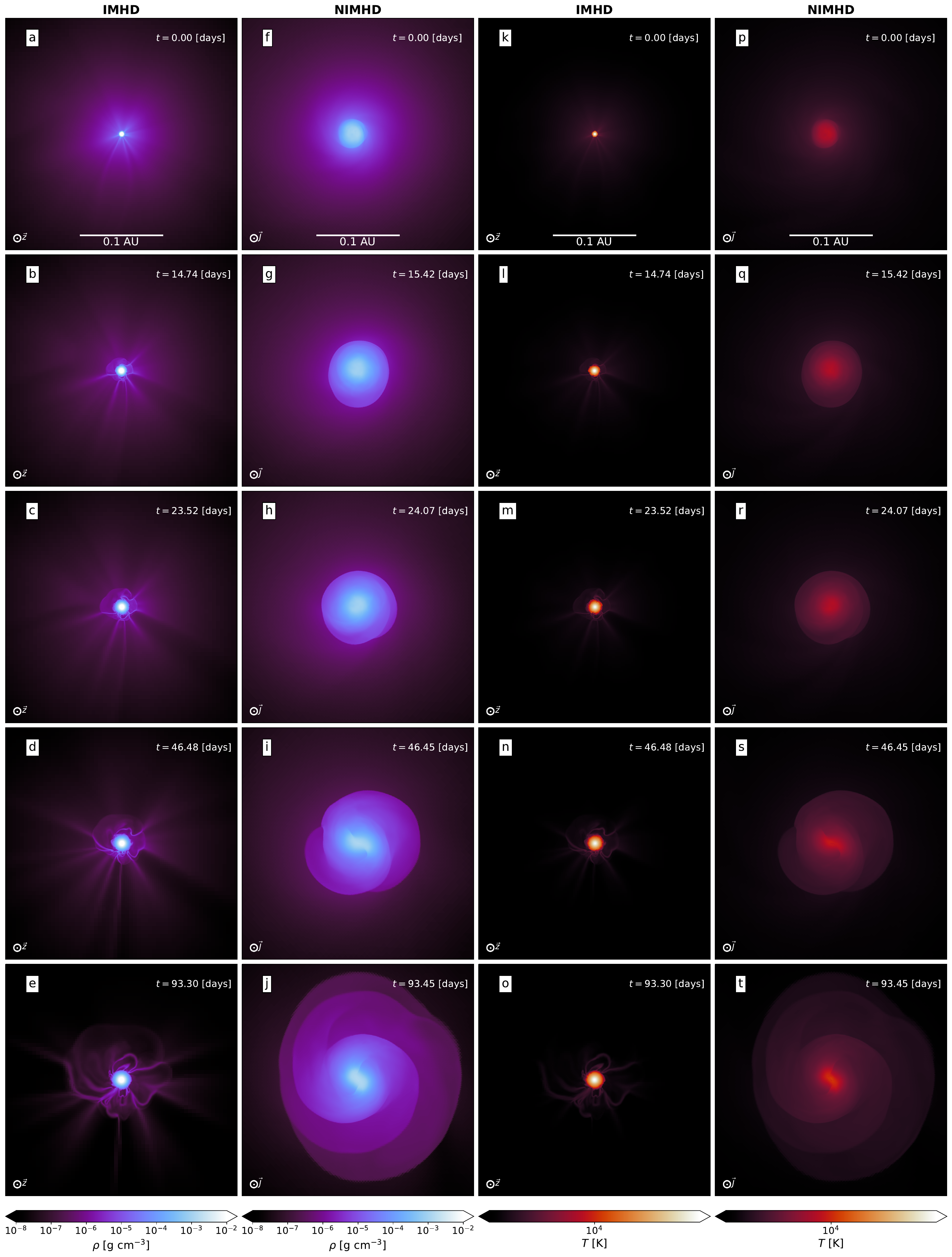} 
\caption[Temperature and density evolution after second collapse.]{A set of slices showing the evolution of the density (first two columns) and temperature (last two columns) for run IMHD (first and third column, panels a-e and k-o) and run NIMHD (second and fourth column, panels f-j and p-t). Each row represents a different time, where $t=0$ corresponds to the moment of protostellar birth. For comparative purposes, the slices are shown at similar times, and the timestamp is written in the top right corner of each panel. The slices are done in the $z$ direction for run IMHD, and along the angular momentum vector for run NIMHD. The scale bars in the first row apply to all other rows as well.}
\label{fig:rhoT}
\end{figure*}

\begin{figure*}[h]
\centering
\includegraphics[scale=.28]{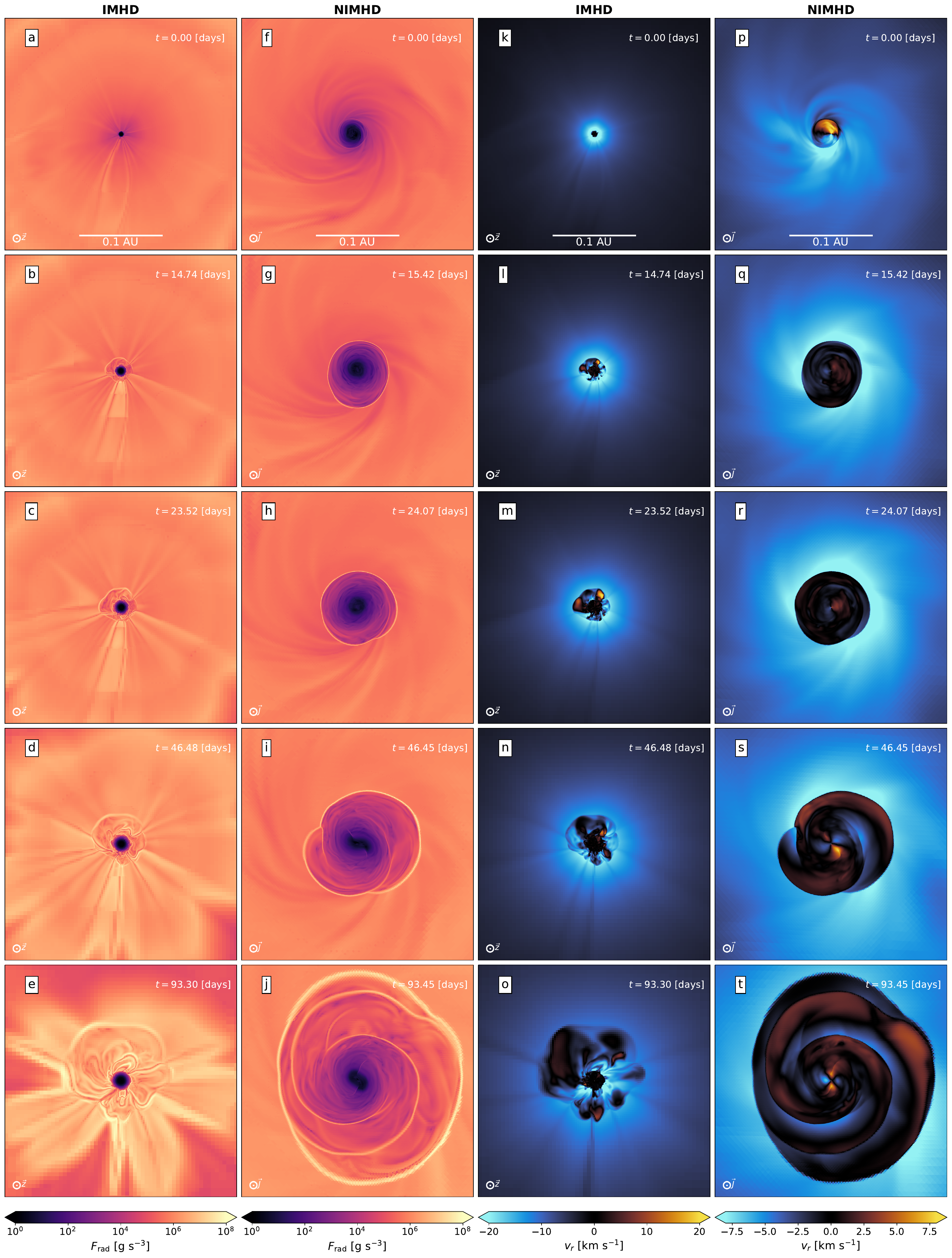} 
\caption[Radiative flux and radial velocity evolution after second collapse.]{Same as \hyperref[fig:rhoT]{Fig.~\ref*{fig:rhoT}}, but this time showing the radiative flux (first two columns) and radial velocity (last two columns).}
\label{fig:fradvr}
\end{figure*}

We now turn to the main focus of our study; the structure of the system following the second gravitational collapse. To this end, we first begin by studying the qualitative structure of the system with the aid of  density, temperature, radiative flux, and radial velocity slices displayed in figures \ref{fig:rhoT} and \ref{fig:fradvr}. The slices are projected along the angular momentum vector of the gas within 0.2 AU for run NIMHD, and in the case of run IMHD, along the $z$ axis since there is no angular momentum left in the gas owing to the efficiency of magnetic braking.
\\
The differences between the resulting protostars are stark. The first row displays the system at protostellar birth, which we define as our $t=0$.\footnote{See \hyperref[appendix:defProtostar]{Appendix \ref*{appendix:defProtostar}} for an overview of how the protostar was defined.} The protostar in run IMHD is more compact than run NIMHD, displaying higher densities and temperatures owing to the lack of centrifugal support against gravity. This causes it to form at a radius of $\approx 0.97\ \mathrm{R_{\odot}}$, whereas in run NIMHD, the centrifugal support flattens the protostar considerably and extends its radius to $\approx 4.8\ \mathrm{R_{\odot}}$. In the weeks following the formation of the protostar in run IMHD (panels b-e and l-o of \hyperref[fig:rhoT]{Fig.~\ref*{fig:rhoT}} and \hyperref[fig:fradvr]{Fig.~\ref*{fig:fradvr}}), its size grows considerably as it accretes material from its surroundings. This is due to the subcritical nature of its accretion shock, which struggles to radiate the incoming accretion energy (see \citealp{ahmad_2023}). In addition, filamentary structures protruding from the stellar surface can be seen growing in spatial extent as time progresses. These are in fact current sheets akin to coronal mass loops, which appear as filaments when visualized in 2D slices.
\\
In the case of run NIMHD, an entirely different evolutionary sequence is witnessed. As time progresses, a disk-like structure surrounding the protostar is formed. This is due to the latter's accumulation of angular momentum, after-which it reaches breakup velocity and material is advected outward.\footnote{Measurements showcasing this are presented in \hyperref[appendix:breakup]{Appendix \ref*{appendix:breakup}}.} This result confirms the findings of \cite{ahmad_2024}, in which no magnetic fields were present. As the disk grows in size and in mass, it exhibits spiral waves which form as a result of gravitational instabilities. These spiral waves carry a significant amount of angular momentum outward and cause increased mass accretion rates onto the protostar.
\\
\\
These figures show the importance of including ambipolar diffusion in our calculations, as it allows for enough angular momentum to survive and hence averts the magnetic braking catastrophe. Below, we proceed to providing a more quantitative analysis of the nascent structures, and since the two runs yielded drastically different results, we define the protostar in two different ways which are presented in \hyperref[appendix:defProtostar]{Appendix \ref*{appendix:defProtostar}}.

\subsubsection{Gas structure and kinematics}

\begin{figure*}[h]
\centering
\includegraphics[scale=.22]{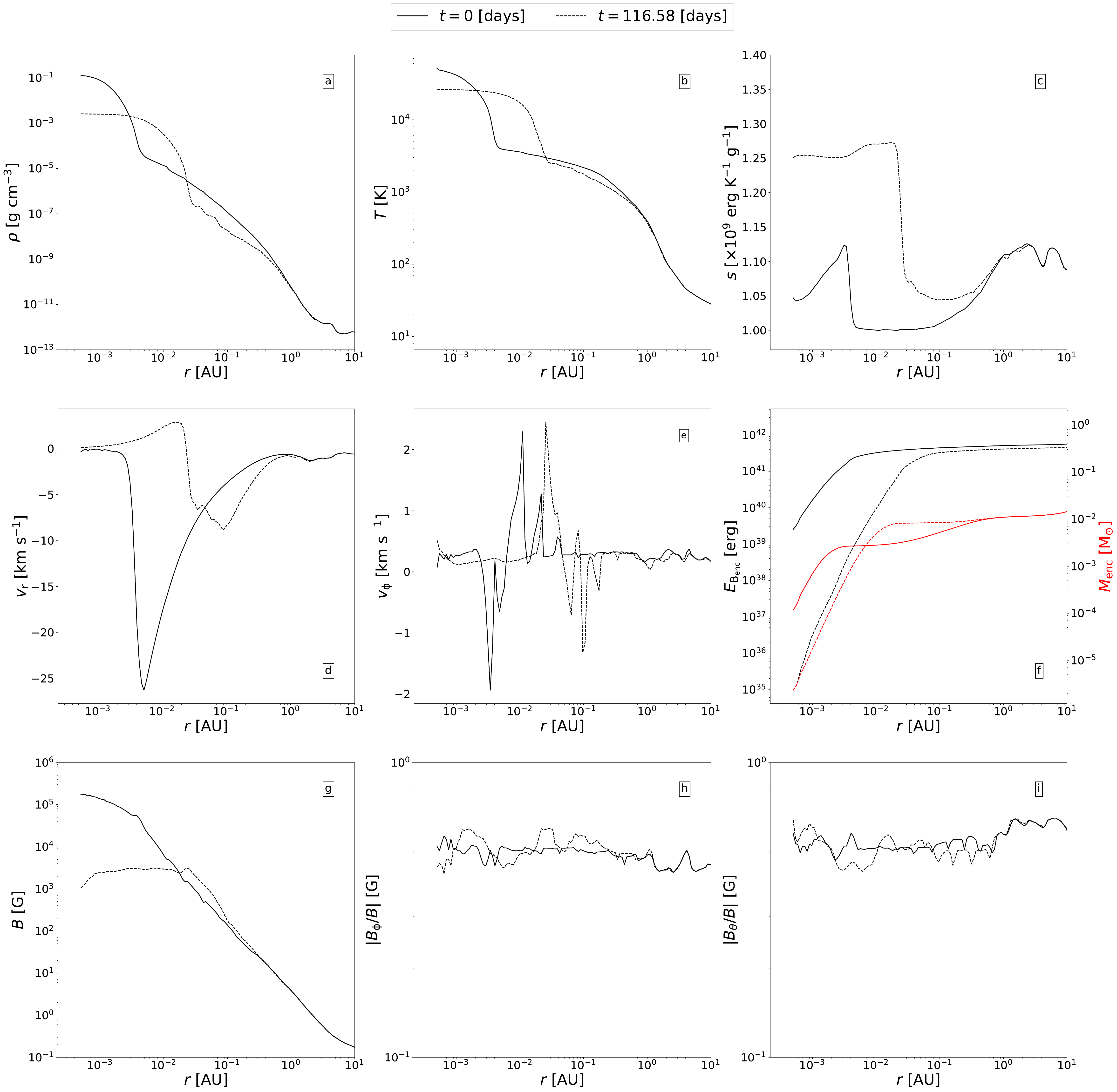} 
\caption[Averaged radial profiles of run IMHD]{A set of measurements of various physical properties of run IMHD at protostellar birth ($t=0$, solid lines) and $t\approx117$ days later (dotted lines). These are averages in spherical bins, and show the gas density (panel a), temperature (panel b), entropy (panel c), radial and azimuthal velocity (panels d and e), magnetic field intensity (panel g), and  the azimuthal and meridional components of the magnetic field, normalized by its magnitude (panels h and i). Panel (f) displays the enclosed magnetic energy (black lines) and the enclosed mass (red lines) as a function of spherical radius.}
\label{fig:imhdprofiles}
\end{figure*}

\begin{figure*}[h]
\centering
\includegraphics[scale=.22]{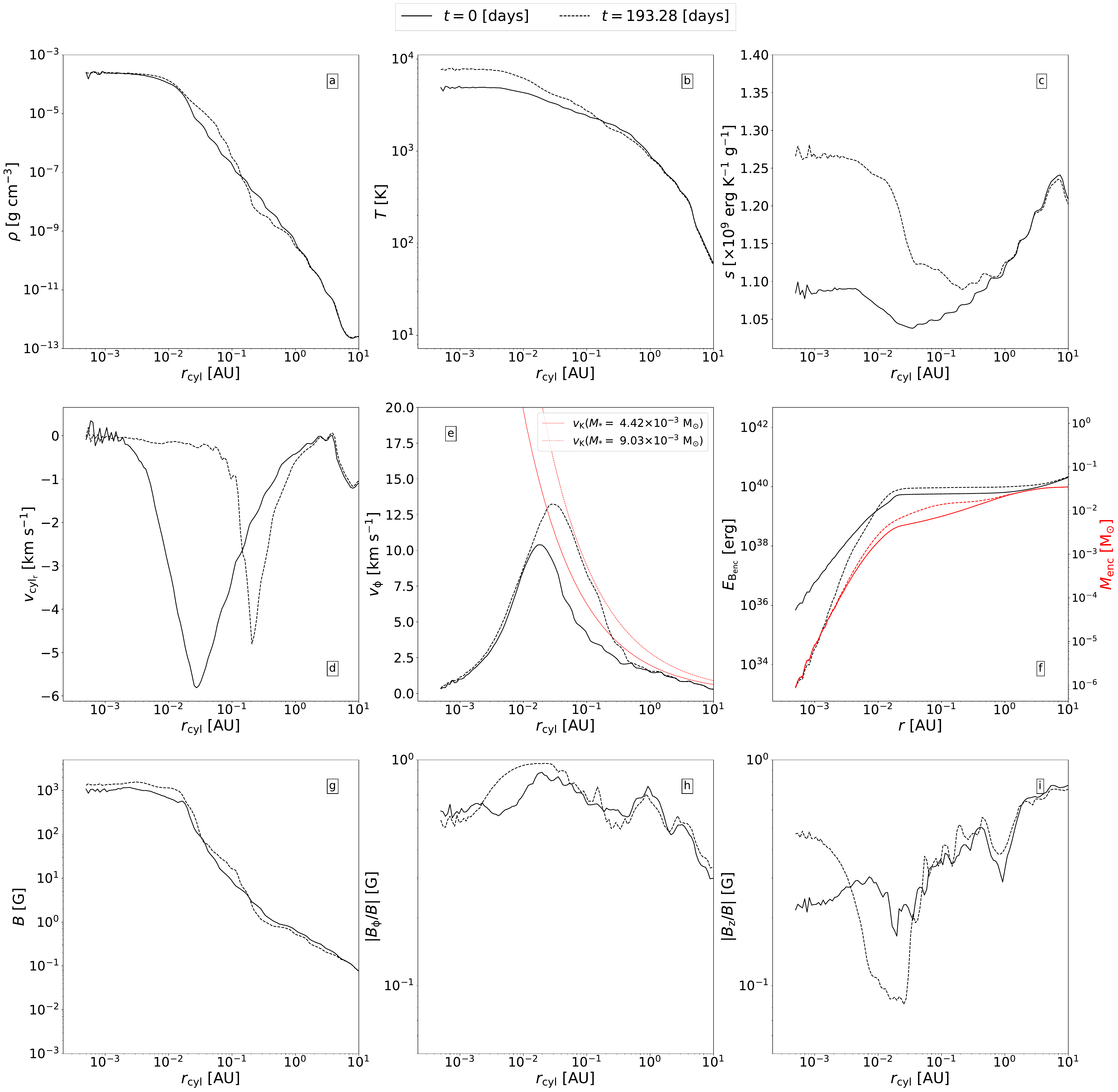} 
\caption[Averaged radial profiles of run NIMHD]{A set of measurements of various physical properties of run NIMHD. These were done in cylindrical radial bins, in which only cells belonging to the midplane, defined as $z\in [-2.5;\ 2.5]\times 10^{-2}$ AU, were used. Solid lines are measurements done at $t=0$ (corresponding to the moment of protostellar birth), and dotted lines are measurements done $\approx 190$ days later. Panel (a), (b), and (c) display respectively the gas density, temperature, and specific entropy. Panels (d) and (e) display the gas' (cylindrical) radial and azimuthal velocity. The red curves in panel (e) display the Keplerian velocity computed with the protostar's mass at a given snapshot. Panels (g), (h), and (i) display the magnetic field strength, its toroidal component, and its vertical component. The toroidal and vertical components are normalized by the total magnetic field strength. Panel (f) displays the enclosed magnetic energy (black lines) and the enclosed mass (red lines) as a function of spherical radius.}
\label{fig:nimhdprofiles}
\end{figure*}

Having ascertained the qualitative structure of the system in the two runs, we proceed to providing a more quantitative comparison between run IMHD and run NIMHD. To this end, we display in \hyperref[fig:imhdprofiles]{Fig.~\ref*{fig:imhdprofiles}} and \hyperref[fig:nimhdprofiles]{Fig.~\ref*{fig:nimhdprofiles}} averages of various physical quantities. In the case of run NIMHD, since the structure we witness is a flattened disk-like structure, these quantities are averaged azimuthaly in cylindrical bins in which only cells in the midplane region are selected. The midplane is defined as the region in which $z\in [-2.5;\ 2.5]\times 10^{-2}$ AU, where the $z$ component is computed along the angular momentum axis of the gas within 0.1 AU. In the case of run IMHD, the measurements are done using the spherical coordinate system since the protostar and the distribution of material around it posses a spherical morphology.
\\
\\
We begin by studying the structure of run IMHD (\hyperref[fig:imhdprofiles]{Fig.~\ref*{fig:imhdprofiles}}). The density profile at protostellar birth, displayed in \hyperref[fig:imhdprofiles]{Fig.~\ref*{fig:imhdprofiles}}~(a) (solid line), shows that the central region of the protostar reaches $\sim 10^{-1}\ \mathrm{g\ cm^{-3}}$.\footnote{This value was shown to be unconverged in \cite{ahmad_2023}, albeit not by a lot and the resolution is such that its numerical outcomes are reliable enough for physical interpretation.} Nearly 117 days later (dotted line), this value drops to $\sim 10^{-3}\ \mathrm{g\ cm^{-3}}$. In both snapshots, a power-law tail follows the central density peak. The sharp discontinuity in the radial velocity profile (\hyperref[fig:imhdprofiles]{Fig.~\ref*{fig:imhdprofiles}}~d) displays the location of the accretion shock, which is $\approx 4\times 10^{-3}$ AU ($\approx 0.86\ \mathrm{R_{\odot}}$) and subsequently moves outward as the protostar expands. The azimuthal velocity curves, shown in \hyperref[fig:imhdprofiles]{Fig.~\ref*{fig:imhdprofiles}}~(e), display the efficiency of magnetic braking in this simulation: nearly no angular momentum survived, as $v_{\mathrm{\phi}}$ alternates between positive and negative values and is mostly a noisy measurement. In \hyperref[fig:imhdprofiles]{Fig.~\ref*{fig:imhdprofiles}}~(c), the specific entropy of the gas\footnote{The specific entropy is obtained through an interpolation of the equation of state table.} is shown. Here, as in \cite{ahmad_2023}, we see that $\mathrm{d}s/\mathrm{d}r > 0$  throughout the protostellar interior, meaning that the protostar is radiatively stable against convective instabilities. However, as accretion still drives turbulence within the protostellar interior \citep{bhandare_2020, ahmad_2023}, the entropy profile at our final snapshot is flattened as a result of the mechanical transport of energy. The entropy profile has also been lifted upwards as a result of the accretion of energy. These results are very similar to the spherically symmetrical RHD run presented in \cite{ahmad_2023}, which again illustrates how efficient magnetic braking is in this run.
\\
\\
In \hyperref[fig:nimhdprofiles]{Fig.~\ref*{fig:nimhdprofiles}} however, we again witness very different results for run NIMHD. Firstly, the density reached in the central regions is two orders of magnitude lower and at $\sim 10^{-3}\ \mathrm{g\ cm^{-3}}$, a value close to the hydro runs presented in \cite{ahmad_2024}. Unlike in run IMHD however, no hydrostatic bounce occurs and the maximum density remains constant as time progresses. The temperature shown in \hyperref[fig:nimhdprofiles]{Fig.~\ref*{fig:nimhdprofiles}}~(b) is also an order of magnitude lower than in run IMHD, and sits close to $5\times 10^{3}$ K, however this increases to $7\times 10^{3}$ K nearly 191 days later, meaning that the protostar is heating up as it accretes material. The cylindrical radial velocity $v_{\mathrm{cyl}}$, displayed in \hyperref[fig:nimhdprofiles]{Fig.~\ref*{fig:nimhdprofiles}}~(d), shows that the protostellar accretion shock is formed at $2\times 10^{-2}$ AU ($\approx 4.3\ \mathrm{R_{\odot}}$), which is nearly five times larger than in run IMHD. This midplane shock front expands outward to $2\times 10^{-1}$ AU at our final snapshot. We emphasize that it is no longer the protostellar accretion shock that is displayed by the discontinuity in $v_{\mathrm{cyl}}$, but rather, that of the newly-formed circumstellar disk in which the protostar is embedded. In \hyperref[fig:nimhdprofiles]{Fig.~\ref*{fig:nimhdprofiles}}~(e), we display the azimuthal velocity curves. Here, we see very clearly that rotational motion exists, as $v_{\phi} > 0$ throughout the radii displayed in the figure. Furthermore, these curves show that the central regions are in solid body rotation, whereas the circumstellar disk exhibits differential rotation. At $t\approx 190$ days, the disk displays a fully Keplerian ($v_{\mathrm{K}}=\sqrt{GM_{*}/r}$) rotation profile (dashed red curve).
\\
Finally, we display in \hyperref[fig:nimhdprofiles]{Fig.~\ref*{fig:nimhdprofiles}}~(c) the specific entropy of the gas. As in run IMHD, the protostar is radiatively stable against convective motion. However, at our final simulation snapshot, we see that it is no longer the case as there exists a region in which $ds/dr<0$. The existence of this region is likely due to the prominent spiral waves within the disk, and the negative entropy gradient means that a convective instability occurs \citep{schwarzschild_1906}, which further contributes to turbulent motion within the disk and hence enhances accretion onto the protostar.
\\
\\
This analysis once again shows the stark differences of both runs; whereas the almost entirely complete absence of angular momentum in run IMHD owing to magnetic braking causes the second collapse to form structures more akin to those produced in spherically symmetrical calculations, the inclusion of ambipolar diffusion allows a considerable amount of angular momentum to survive and hence form a rotationally supported disk surrounding the protostar. In this sense, run NIMHD is more related to hydrodynamical runs than to run IMHD, and thus should be quantitatively and qualitatively compared as such.

\subsection{Magnetic field structure}
\label{sec:magfieldstruct}

\begin{figure*}[h]
\centering
\includegraphics[scale=.25]{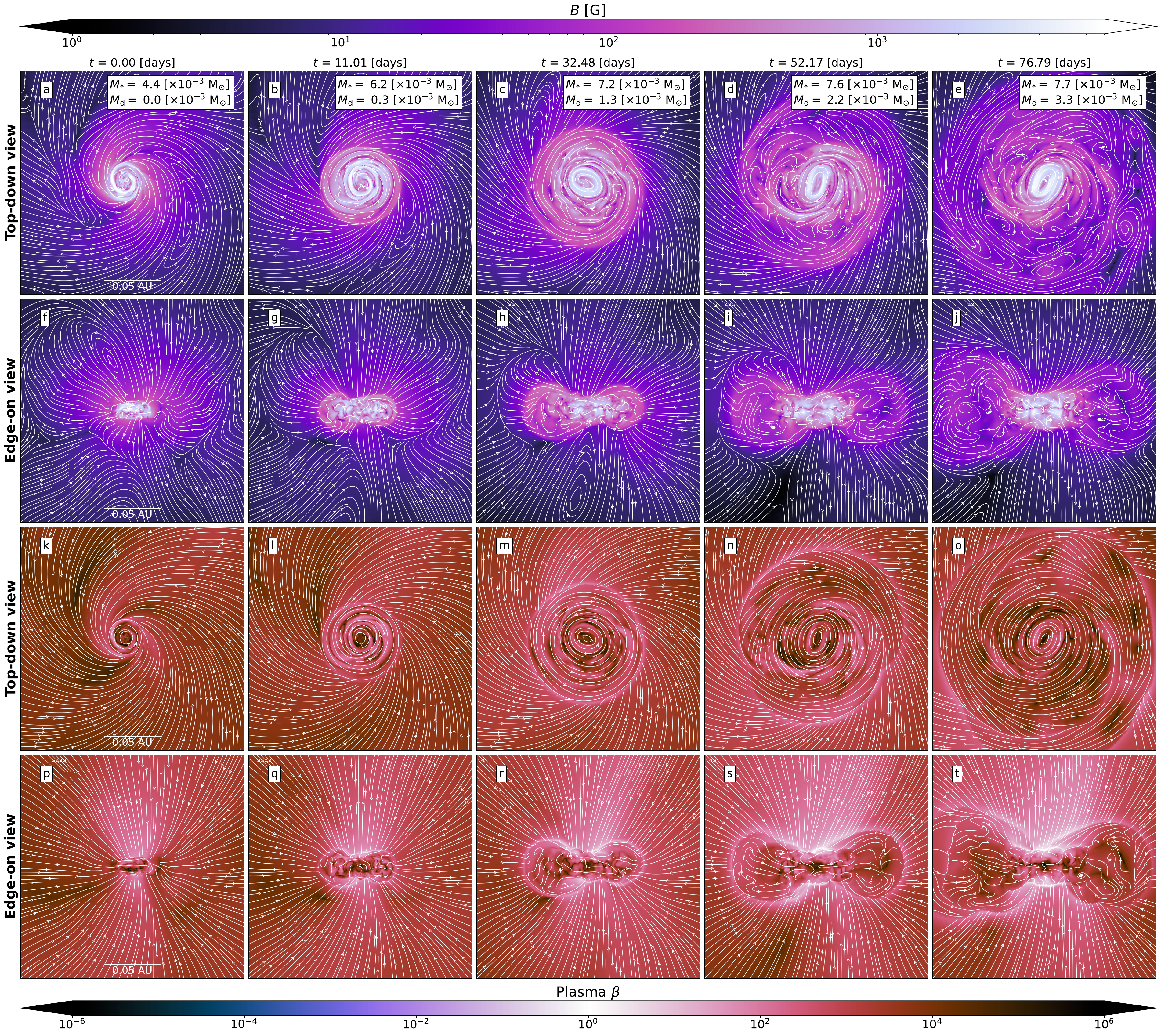} 
\caption[Magnetic field strength and morphology in the resistive case]{The magnetic field of run NIMHD: a set of slices displaying the magnetic field strength (top two rows) and plasma $\beta$ (bottom two rows) in a top-down (first and third rows) and edge-on (second and fourth rows) view. The white curves in the first two rows (panels a-j) are magnetic field streamlines, whereas in the last two rows (panels k-t) they are velocity vector field streamlines. Each column represents a different time, where $t=0$ corresponds to the moment of protostellar birth. The mass of the protostar and its circumstellar disk is displayed in the top right corner of panels a-e. The scale bars in the first column apply to all other columns as well.}
\label{fig:nimhdbfield}
\end{figure*}

\begin{figure*}[h]
\centering
\includegraphics[scale=.25]{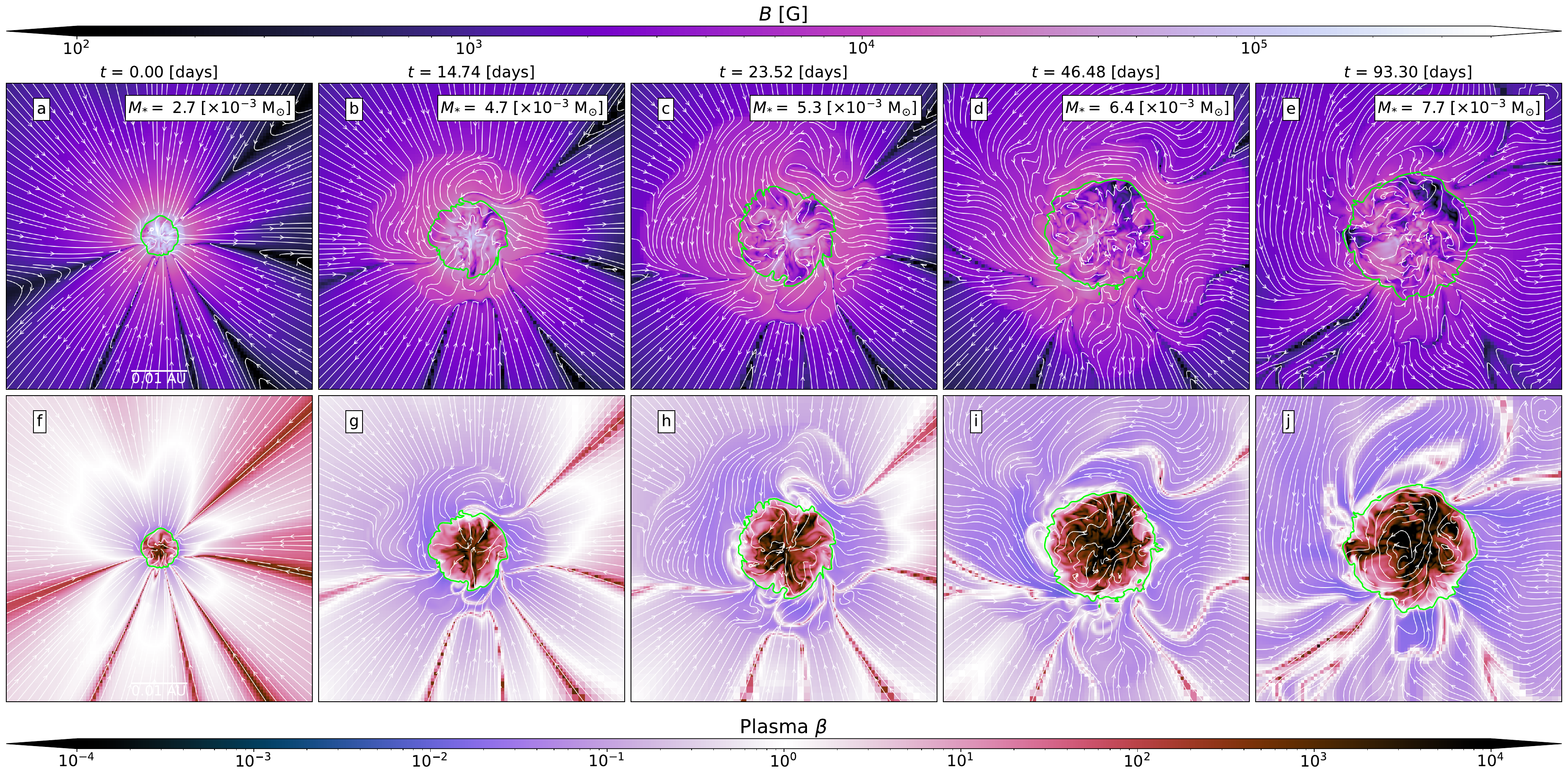} 
\caption[Magnetic field strength and morphology in the ideal MHD limit]{Same as \hyperref[fig:nimhdbfield]{Fig.~\ref*{fig:nimhdbfield}}, but for run IMHD. Since there is virtually no rotation in this run, the slices are done along the $z$ axis only. The lime-colored contour designates the stellar surface.}
\label{fig:imhdbfield}
\end{figure*}

In this section, we describe the structure and morphology of the magnetic field within and in the close vicinity of the protostar. To this end, we display in \hyperref[fig:nimhdbfield]{Fig.~\ref*{fig:nimhdbfield}} and \hyperref[fig:imhdbfield]{Fig.~\ref*{fig:imhdbfield}} slices showing the magnetic field strength and plasma $\beta$ ($=8\pi P/B^{2}$). In \hyperref[fig:nimhdbfield]{Fig.~\ref*{fig:nimhdbfield}}, the slices are shown in a top-down (panels a-e and k-o) and edge-on (panels f-j and p-t) view, whereas the absence of rotation in run IMHD renders any such double-visualization useless, and we may visualize these quantities along the $z$ axis only. We will also leverage the information available in \hyperref[fig:imhdprofiles]{Fig.~\ref*{fig:imhdprofiles}} and \hyperref[fig:nimhdprofiles]{Fig.~\ref*{fig:nimhdprofiles}}.
\\
\\
The magnetic field streamlines of run NIMHD are displayed in panels (a-j) of \hyperref[fig:nimhdbfield]{Fig.~\ref*{fig:nimhdbfield}}. In the top-down view (\hyperref[fig:nimhdbfield]{Fig.~\ref*{fig:nimhdbfield}}, panels a-e), we see the magnetic field lines being dragged by the nascent protostar. At the temperature-density regime displayed here, the ideal MHD limit is recovered as all dust grains are sublimated and ambipolar diffusion is no longer at play. In addition to this, the gas begins to ionize, which further enhances the magnetic field's coupling to it. In addition, the plasma $\beta$ values displayed in the last two rows indicates that thermal pressure support far outweighs magnetic pressure support, meaning that it is the fluid that dictates the behavior of the magnetic field. As such, the rotation of the newly-formed protostar and circumstellar disk causes a significant build-up of the toroidal component of the magnetic field, as the lines are twisted and tangled to an extreme degree by the dynamical second collapse. Along the disk midplane, there also seems to be a significant amount of turbulent magnetic eddies, which appear most prominent at later times. These are likely formed as a result of the emergence of spiral waves (see Figs. \hyperref[fig:rhoT]{\ref*{fig:rhoT}} and \hyperref[fig:fradvr]{\ref*{fig:fradvr}}), which create significant turbulent motion within the disk. In essence, the turbulent eddies show that the magnetic field is at places confined within a tube-like structure which crosses the disk midplane, and hence showcases a significant poloidal component within the disk. We also note the spiral structure of the magnetic field intensity within the star-disk system. In \cite{wurster_2022}, it is claimed that the Hall effect is responsible for the creation of this spiral structure, however our results show here that the Hall effect is not necessary to form it. We show in \hyperref[appendix:GI]{Appendix \ref*{appendix:GI}} that the circumstellar disk is marginally stable against gravitational collapse, with the classical Toomre $Q$ parameter hovering around unity.
\\
Interestingly, in the edge-on view of panels (h-j) of \hyperref[fig:nimhdbfield]{Fig.~\ref*{fig:nimhdbfield}}, we see what appears to be a dipolar field in the western half of the star-disk system, where magnetic field streamlines originating from the southern pole of the protostar loop back into its northern pole, however we are unsure as to why the feature appears only in the western half.\footnote{We believe this feature is likely transient, as it is not as evident at later times.} Outside the star-disk system, the magnetic field lines are mostly vertical and they thread the two bodies, showcasing the poloidal nature of the magnetic field in these regions. The plasma $\beta$ decreases in the polar regions over time due to the depletion of material in these regions as the second collapse proceeds \citep{ahmad_2024}, which in turn causes a reduction in thermal pressure support. The disk's surface also appears to have a plasma $\beta\approx 1$, and the velocity vector field streamlines indicate that material is advected toward the protostar from the upper layers of the disk, as reported previously in the MHD run of \cite{lee_2021} and the hydro runs of \cite{ahmad_2024}. \hyperref[appendix:massflux]{Appendix \ref*{appendix:massflux}} presents a quantitative measurement of the directional mass flux, which shows that the upper layers of the disk transport a similar amount of material towards the interior as the main body of the disk. Despite this, we see no outflow or high velocity jet developing, as the velocity vector field streamlines in panels (p-t) of \hyperref[fig:nimhdbfield]{Fig.~\ref*{fig:nimhdbfield}} are pointing towards the protostar, thus indicating infall. Any such outflows are likely to occur at much later times, when the polar reservoir of gas is significantly depleted and the plasma $\beta$ in these regions drop to very small values. This once again confirms the results of \cite{wurster_2020}, which found that turbulence in the initial dense cloud core significantly delays the onset of jets and outflows. This is likely due to the absence of coherent magnetic field lines, which significantly hinders the onset of jets and outflows. In addition, \cite{vaytet_2018} also reports the absence of jets or outflows at protostellar scales despite the absence of initial turbulence in their progenitor dense core, which is likely due to the fact that the toroidal component of the magnetic field has yet to reach the strength needed to trigger the magneto-centrifugal mechanism \citep{lynden_1996, lovelace_2002}. Finally, \cite{machida_2014} has shown a sensitivity of the presence of jets and outflows with regards to the large scale initial conditions where they struggled to recover the latter when using uniform density progenitor cores (as is used in \cite{vaytet_2018} and the present study).
\\
With regards to the spatial distribution of the magnetic field within the star-disk system, we unsurprisingly see that the central region containing the protostar has the strongest field strength, reaching $\approx 5\times 10^{3}$ G. In accordance with \cite{wurster_2022}'s higher resolution runs, we witness spiral structures in magnetic field strength throughout the star-disk system.
\\
In panels (h) and (i) of \hyperref[fig:nimhdprofiles]{Fig.~\ref*{fig:nimhdprofiles}}, quantitative measurements of $B_{\phi}$ and $B_{\mathrm{z}}$ are provided. The cylindrical radial velocity displayed in \hyperref[fig:nimhdprofiles]{Fig.~\ref*{fig:nimhdprofiles}} (d) allows one to locate the accretion shock, which manifests itself as a strong discontinuity.\footnote{The asymmetrical distribution of matter in the equatorial regions caused by the disk's eccentricity dilutes the azimuthal average and causes $v_{\mathrm{cyl}}$ to diffuse.} Firstly, at $t=0$, the toroidal component is the dominant one within the protostar. However, at $t\approx 191$ days, the vertical component is significantly built up and it becomes stronger than its toroidal counterpart. However this appears to be transient, as the vertical component within the protostar seems to be oscillating. At larger radii (i.e., within the circumstellar disk), the opposite occurs: we see a build-up of the toroidal component of the magnetic field whereas the poloidal component is significantly reduced. In \hyperref[fig:nimhdprofiles]{Fig.~\ref*{fig:nimhdprofiles}} (f), the black lines display the enclosed magnetic energy within the (spherical) radius $r$, which is computed as:
\begin{equation}
E_{B_{\mathrm{enc}}} = \frac{1}{2}\int_{0}^{r}B^{2}r^{2}dr\ .
\end{equation}
We see that the innermost regions of the system lose magnetic energy over time. In these regions, the gas recovers the ideal MHD limit and flux freezing holds, with $B\propto \rho^{2/3}$. Since the density within said regions remains somewhat constant, their loss of magnetic energy is due to an outward advection of material, as the protostar exceeds breakup velocity and begins shedding its surface material \citep{ahmad_2024}.
\\
\\
We now turn to describing the magnetic field structure of run IMHD (\hyperref[fig:imhdbfield]{Fig.~\ref*{fig:imhdbfield}}). Here, at $t=0$ (panels (a) and (f) of \hyperref[fig:imhdbfield]{Fig.~\ref*{fig:imhdbfield}}), we see an extreme pinching of the magnetic field lines as a result of the second gravitational collapse. In addition, the field lines outside the protostar (lime contour) are almost entirely radial, with virtually no toroidal component present. However, as in \cite{ahmad_2023}, we find strong turbulent motion within the protostar.\footnote{This is better seen in later times displayed in the figure.} This causes all magnetic field components within the protostar to reach a similar strength, a fact that is particularly evident in panels (h) and (i) of \hyperref[fig:imhdprofiles]{Fig.~\ref*{fig:imhdprofiles}}.
\\
One final result we would like to report is in regards to the slices shown in panels (f-j) of \hyperref[fig:imhdbfield]{Fig.~\ref*{fig:imhdbfield}}, displaying the plasma $\beta$ of the gas. In \cite{vaytet_2018}, it is reported that the protostar formed under the ideal MHD approximation is a magnetically supported object. However, we show here that all gas downstream of the protostellar accretion shock (lime contour) has a plasma $\beta \approx 1$ or $\gg 1$. This means that the protostar is an entirely thermally supported body.\footnote{We believe that the interpretation in \cite{vaytet_2018} is an oversight owing to the fact that they looked at 2D histograms of plasma $\beta$, rather than slices.} In addition, we have overlayed on these slices the velocity vector field streamlines, which show that no outflow or jet is being launched by the protostar. This is to be expected given the immensely unstructured nature of the magnetic field in this run, which exhibits no coherent toroidal component owing to the lack of rotational motion, and can thus no longer drive an outflow through the magneto-centrifugal mechanism \citep{blandford_1982, ouyed_1997}. This is in agreement with \cite{wurster_2020}.



\subsection{Disk expansion: comparison with the hydro case}
\label{sec:diskexpansion}

\begin{figure*}[h]
\centering
\includegraphics[scale=.35]{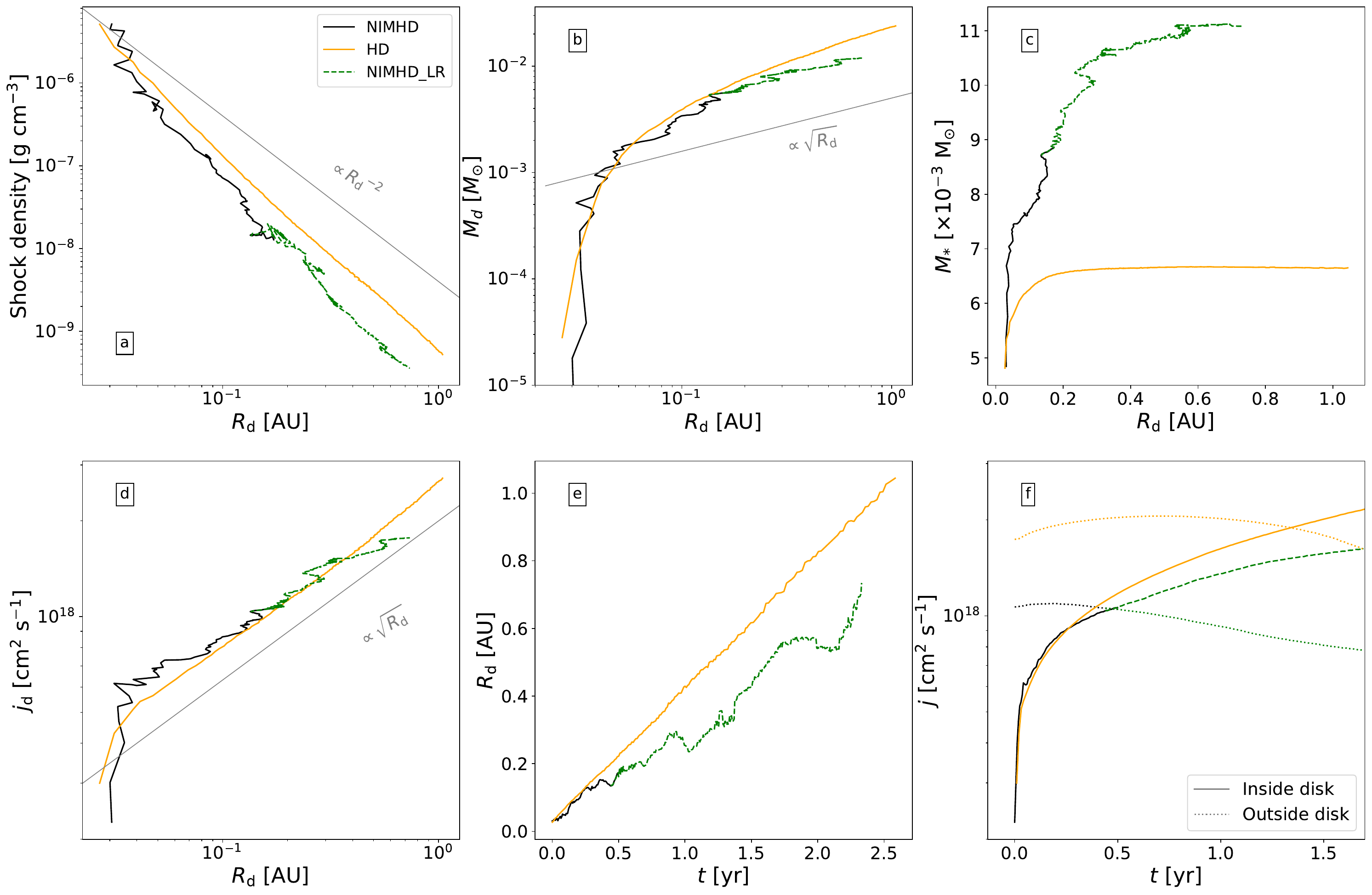} 
\caption[Comparing the disk evolution in the hydro and MHD case]{Temporal evolution of the circumstellar disk of run NIMHD (black curves), compared with its hydro counterpart (orange curves, taken from \cite{ahmad_2024}). The green curves are a zoom-out branched from run NIMHD and run at a lower resolution with $\ell_\mathrm{max}=24$ (see \hyperref[sec:zoomout]{Sec. \ref*{sec:zoomout}}). Panels (a), (b), (c), and (d) display as a function of the disk's equatorial radius $R_{\mathrm{d}}$ respectively the density measured at the disk's equatorial shock front (obtained through ray-tracing), the mass of the disk, the mass of the protostar, and the disk's specific angular momentum. Panel (e) displays $R_{\mathrm{d}}$ as a function of time, where $t=0$ marks the moment of birth of the disk. Panel (f) displays the specific angular momentum of the gas within 1 AU found inside the disk (solid and dashed lines) and outside the disk (dotted lines) as a function of time since the birth of the disk.}
\label{fig:diskevol}
\end{figure*}

\begin{figure*}[h]
\centering
\includegraphics[scale=.45]{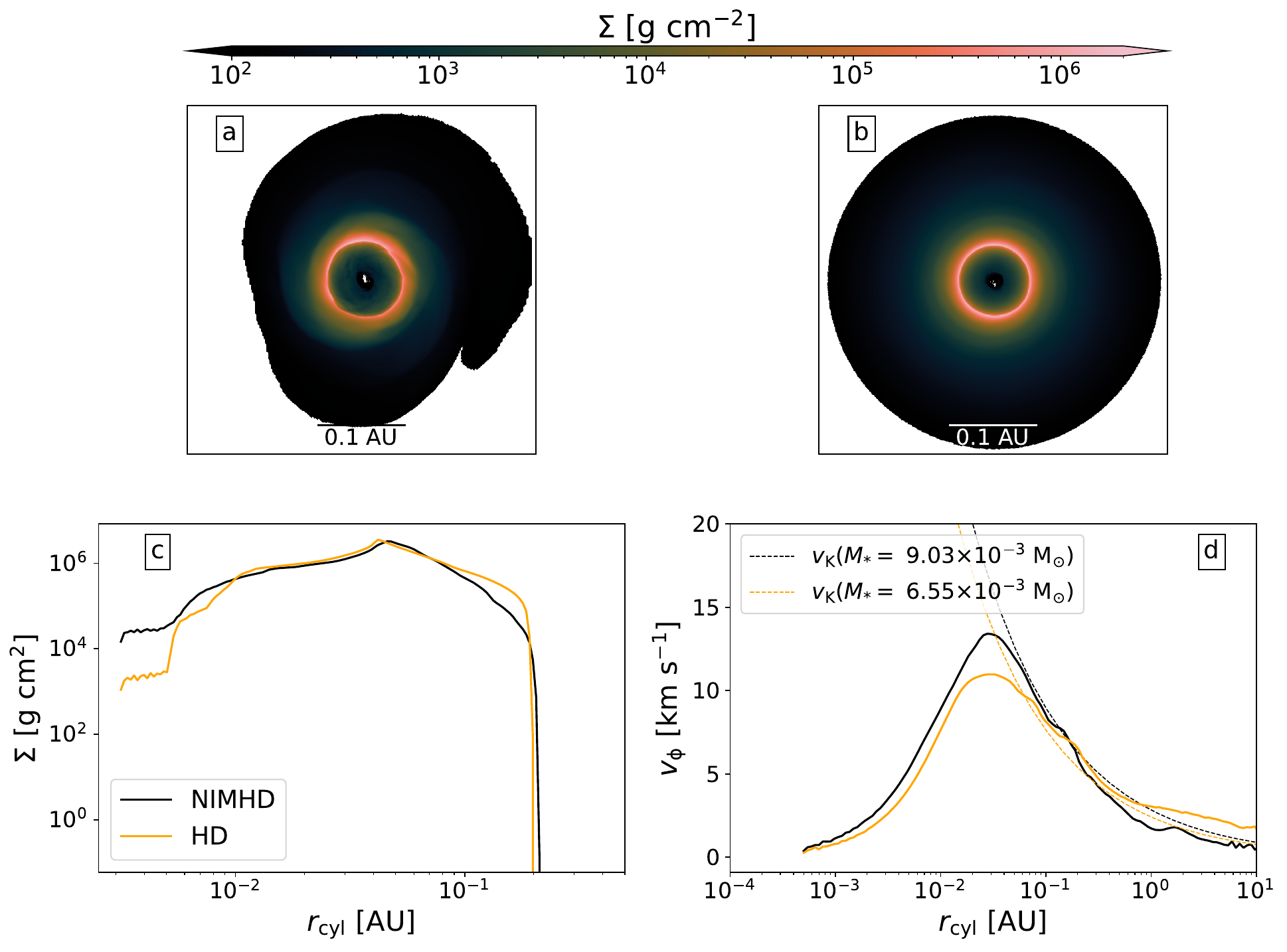} 
\caption[Structural and kinematic comparison between run NIMHD and HD]{A structural and kinematic comparison between run NIMHD and HD (resp. black and orange curves in panels c and d). Panel (a) displays column density maps for run NIMHD at our final simulation snapshot, which is $\approx 190$ days after protostellar birth. Panel (b) displays the equivalent map for run HD, at a moment in time where its radius is comparable to that of run NIMHD ($\approx 0.2$ AU). Only cells belonging to the disk were used in the making of these maps. The second row displays the radial profiles of column density (panel c) and azimuthal veocity (panel d). The dashed lines in panel (d) display the Keplerian velocity computed with the protostar's mass.}
\label{fig:nimhdhydrocomp}
\end{figure*}

We now turn to providing a quantitative description of the evolution of the circumstellar disk over time. Since its structure appears to be qualitatively similar to the hydro case (a large and highly flared disk), we will compare it to that obtained in run G2 of \cite{ahmad_2024}, whose initial conditions and numerical setup are the same (notwithstanding the absence of magnetic fields). In addition, since \hyperref[fig:nimhdbfield]{Fig.~\ref*{fig:nimhdbfield}} has shown that the plasma $\beta \gg 1$ within the disk, we expect a similar evolution to the hydro case but with a notable increase in torquing owing to the strong magnetic field strength within the disk ($\sim [10^{2}-10^{3}]$~G) and in the envelope ($\sim 10^2$ G). To this end, we display in \hyperref[fig:diskevol]{Fig.~\ref*{fig:diskevol}} the mass, radius, specific angular momentum, and density at the equatorial shock front of the circumstellar disk in run NIMHD (black curves) and in the hydro run of \cite{ahmad_2024} (orange curves, hereafter run HD). We also leverage the information provided in \hyperref[fig:nimhdhydrocomp]{Fig.~\ref*{fig:nimhdhydrocomp}}, which displays the column density maps of runs NIMHD and HD (resp. panels (a) and (b) of \hyperref[fig:nimhdhydrocomp]{Fig.~\ref*{fig:nimhdhydrocomp}}) and their corresponding radial profiles ( \hyperref[fig:nimhdhydrocomp]{Fig.~\ref*{fig:nimhdhydrocomp}}~c). The velocity profiles are shown in  \hyperref[fig:nimhdhydrocomp]{Fig.~\ref*{fig:nimhdhydrocomp}}~(d).
\\
Additionally, since it is of interest to advance the simulation in time, we have branched run NIMHD at $\approx 0.4$ years following protostellar birth and run a parallel simulation with a reduced $\ell_{\mathrm{max}}=24$, which significantly alleviates the computational cost of the simulation. The properties of the disk in this run, labeled "NIMHD\_LR", are shown in the green curves of \hyperref[fig:diskevol]{Fig.~\ref*{fig:diskevol}}. The overlap between the black and green curves shows that its results are realistic enough for physical interpretations.
\\
\\
We first begin by studying the temporal evolution of the disk's radius with respect to time (\hyperref[fig:diskevol]{Fig.~\ref*{fig:diskevol}}~e). We note the fact that although the evolution in the initial 0.3 years are identical in the HD and NIMHD runs, they later diverge as run NIMHD exhibits a slower disk growth in time owing to strong torque mechanisms. Nevertheless, the plot displaying disk's mass with respect to its radius (\hyperref[fig:diskevol]{Fig.~\ref*{fig:diskevol}}~b) shows that the HD and NIMHD runs exhibit a similar evolutionary trend, although run NIMHD's curves appears more oscillatory than that of run HD. These oscillations are caused by strong spiral waves within the disk in run NIMHD, which carry a significant amount of material inwards and in doing so, also reduce the mass of the disk when compared to run HD. Indeed, at a given disk radius, run NIMHD displays a smaller mass than run HD. These spiral waves are likely caused by magnetic torques, which reduce the gas's centrifugal support against gravity, and they warp the disk \citep{lai_2003, tomida_2013}, as can be seen in \hyperref[fig:nimhdhydrocomp]{Fig.~\ref*{fig:nimhdhydrocomp}}~(a). Indeed, when measuring the mass-weighted mean magnetic field strength within a radius of 10 AU, the toroidal component outweighs all others by an order of magnitude ($\approx 270$ G, compared to $\approx 60$ G for the cylindrical radial component and $\approx 25$ G for the vertical field). This means that the magnetic field is mostly parallel to the disk midplane, which has been shown to cause prominent spiral waves to develop \citep{joos_2012, li_2013, hennebelle_2020}. In contrast, the hydro disk (\hyperref[fig:nimhdhydrocomp]{Fig.~\ref*{fig:nimhdhydrocomp}}~b) remains circular. This causes higher column densities in the outer regions of the disk in run HD than in run NIMHD (\hyperref[fig:nimhdhydrocomp]{Fig.~\ref*{fig:nimhdhydrocomp}}~c). Finally, the higher protostellar mass causes faster rotation in the innermost regions of the disk in run NIMHD (\hyperref[fig:nimhdhydrocomp]{Fig.~\ref*{fig:nimhdhydrocomp}}~d), and its velocity profile closely approaches the Keplerian profile (dashed line), whereas in run HD the velocity profile is super-Keplerian owing to the disk's self-gravity. The disk is slightly sub-Keplerian in run NIMHD owing to thermal pressure support. Magnetic pressure support has a negligible effect on $v_{\mathrm{\phi}}$ since $\beta \ll 1$.
\\
The prominence of magnetic braking is further displayed in \hyperref[fig:diskevol]{Fig.~\ref*{fig:diskevol}}~(c), which shows the protostellar mass as a function of disk radius. Here, we see that although run HD quickly reaches a plateau in disk mass, run NIMHD shows a rapidly growing protostar. Note that the specific angular momentum of the disk, shown in \hyperref[fig:diskevol]{Fig.~\ref*{fig:diskevol}}~(d), is the same in both runs and scales as $\sqrt{R_{\mathrm{d}}}$.\footnote{This scaling is a consequence of the disk's Keplerian velocity profile.} In \hyperref[fig:diskevol]{Fig.~\ref*{fig:diskevol}}~(f), we show the specific angular momentum of the gas within 1 AU, computed both outside of the disk (dotted lines) and within it (solid and dashed lines). This figure shows that the disk in run NIMHD is accreting from a reservoir of gas containing a smaller amount of angular momentum than run HD, which shows that magnetic braking occurred before the gas was accreted onto the disk.
\\
The reduced mass of the disk in run NIMHD when compared to run HD also manifests itself in a reduced disk density. More specifically, when measuring the density at the disk's equatorial shock front ($\rho_{\mathrm{acc}}$, \hyperref[fig:diskevol]{Fig.~\ref*{fig:diskevol}}~a), we see that it is consistently lower than in run HD. Although we could not integrate the calculations to the point where the disk reaches the commonly used sink accretion radius of 1 AU, we present in \hyperref[appendix:evenlowerres]{Appendix \ref*{appendix:evenlowerres}} the results of a run branched out of NIMHD\_LR (labeled NIMHD\_LR\_2), whose lower resolution allowed for the calculation to be pushed longer out in time. It predicts $\rho_{\mathrm{acc}}(\mathrm{1\ AU})\approx 2\times 10^{-10}\ \mathrm{g\ cm^{-3}}$, a value that is a factor $\approx 3$ lower than that predicted by the hydrodynamical run ($\approx 5.93\times 10^{-10}\ \mathrm{g\ cm^{-3}}$).

\subsection{Eccentricity measurements}

\begin{figure}[h]
\centering
\includegraphics[scale=.45]{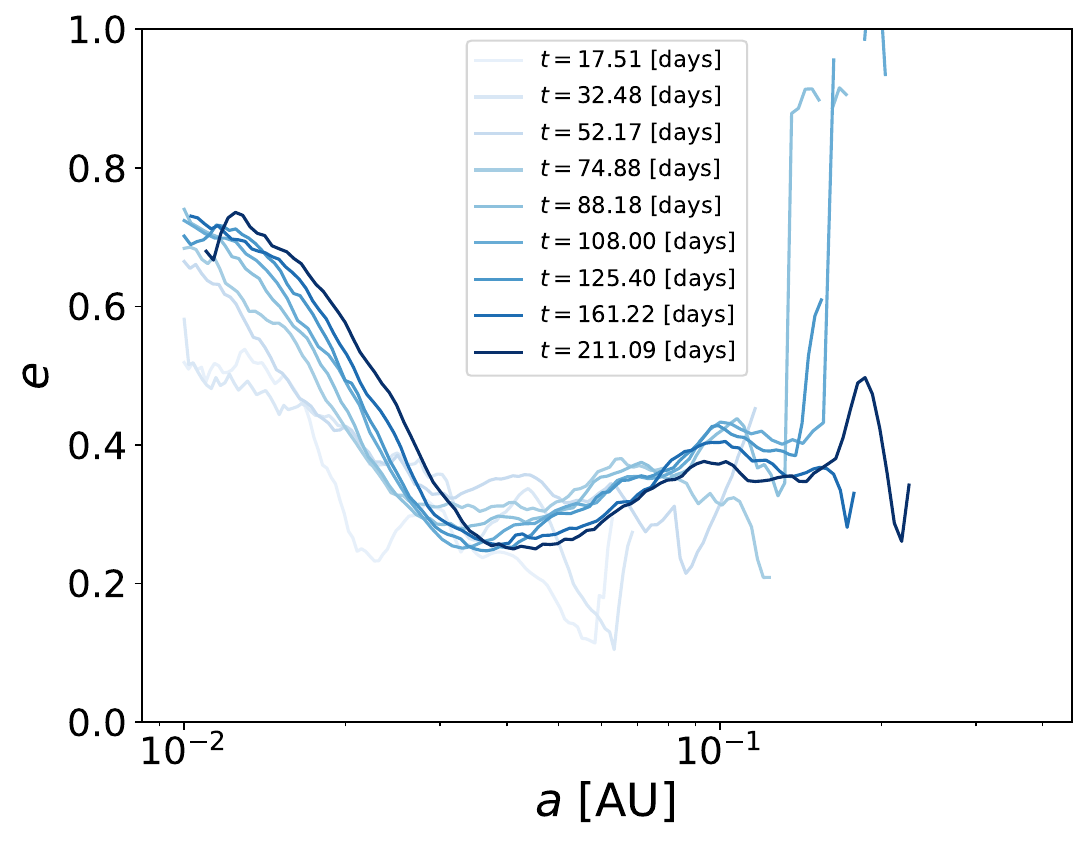} 
\caption{Mass-weighted average of the magnitude of the eccentricity vector as a function of semimajor axis (see Eq. \ref{eq:e} and \ref{eq:a}) in run NIMHD. Each curve corresponds to a different time, where $t=0$ corresponds to the epoch of protostellar birth.}
\label{fig:eccentricityNIMHD}
\end{figure}

Recently, \cite{commercon_2024} has shown that disks born out of gravitational collapses exhibit strong eccentricity. In the present study, run NIMHD has yielded a disk that also appears to show eccentric motion, as is most prominently seen in the asymmetrical flaring of the disk in panels (p-t) of \hyperref[fig:nimhdbfield]{Fig. \ref*{fig:nimhdbfield}}. As such, we provide quantitative measurements of the eccentricity of the circumstellar disk in run NIMHD. In order to do so, we compute both the eccentricity vector $\Vec{e}$ and semimajor axis $a$, defined as \citep{commercon_2024}
\begin{equation}
    \label{eq:e}
    \Vec{e} = \frac{\vec{v}\times \Vec{j}}{GM_{\mathrm{enc}}(r)}-\frac{\Vec{r}}{r},
\end{equation}
\begin{equation}
    \label{eq:a}
    a = \frac{rGM_{\mathrm{enc}}(r)}{2GM_{\mathrm{enc}}(r)-v^{2}r},
\end{equation}
where $\Vec{j}$ is the specific angular momentum of the gas and $\Vec{r}$ is the position vector. Since the disk mass is greater than or comparable to the protostellar mass (see panels (b) and (c) of \hyperref[fig:diskevol]{Fig. \ref*{fig:diskevol}}), we compute $e$ and $a$ using $M_{\mathrm{enc}}(r)$, the enclosed mass within $r$:
\begin{equation}
    M_{\mathrm{enc}} = 4\pi\int_{0}^{r}\rho r^{2}dr.
\end{equation}
The resulting measurements of the magnitude of the eccentricity vector $e=|\Vec{e}|$ are presented in \hyperref[fig:eccentricityNIMHD]{Fig. \ref*{fig:eccentricityNIMHD}}, which displays the mass-weighted orbital averages of $e$. We see in this figure strong eccentric motions within the disk, with $e$ reaching values that consistently exceed 0.2. This shows that similarly to larger scale disks forming around sink particles (e.g., as in \cite{lee_2021, commercon_2024}), the circumstellar disk forming as a result of the rotational breakup of the protostar is also eccentric. The strong gradient in $e$ in the outer regions of the disk, seen particularly at later times, are caused by strong spiral motions within the disk.

\section{Discussions} \label{section:discussions}

The results presented here, most notably those of run NIMHD, are noteworthy for a number of outstanding issues in stellar formation theory. We discuss their implications in this wider context below.

\subsection{The angular momentum problem}

In \cite{ahmad_2024}, we reported on the birth of the circumstellar disk as a result of the breakup of the protostar and the subsequent vigorous radial expansion of the disk in time. Run NIMHD has confirmed that such a phenomenon is reproduced even in the presence of magnetic fields, provided that magnetic resistivities are accounted for. In the literature, \cite{machida_2011b, vaytet_2018, wurster_2020, machida_2019, bhandare_2024} have also reported on the birth of a circumstellar disk that rapidly expands to larger radii. \cite{machida_2007} also reported that their non-ideal MHD calculations produced rapidly rotating protostars, and stressed that the calculations must be run further out in time to properly model the angular momentum evolution of prestellar objects.
\\
What these simulations seem to show is that a paradigm shift is required in our understanding of the angular momentum problem. Indeed, long has it been implicitly implied in stellar formation theory that angular momentum must be lost \textit{during} the collapse so as to prevent the protostar from ever reaching breakup velocity \citep{bodenheimer_1995}. What our comparison between run IMHD and NIMHD shows is that should angular momentum transport by magnetic fields be so efficient so as to prevent the protostar from ever reaching breakup velocity, then no circumstellar disk forms: it is the very fact that parcels of fluid within the protostar achieve rotational breakup that allows for circumstellar disks to form in our simulations.\footnote{See \hyperref[appendix:breakup]{Appendix \ref*{appendix:breakup}}.} As such, whatever angular momentum transport process is responsible for spinning down the protostar to the $\sim 10-15\%$ of breakup velocity as is observed in YSOs \citep{rebull_2004, herbst_2007}, it is acting on longer timescales than the free-fall time of the cloud.
\\
The vigorous radial expansion of the disk may even have left its mark on the meteoric record of the solar system, as high-temperature condensates in the form of calcium-aluminum inclusions and amoeboid olivine aggregates show evidence of rapid outward transport \citep{morbidelli_2024}.

\subsection{The magnetic flux problem}

As mentioned previously, YSOs are consistently found to have $\sim$ kG magnetic field strengths, with the earliest measurements being those of Class I sources. It is as of yet unclear what the origin of these magnetic fields is; whether they are fossil fields that are maintained for $\sim 10^{5}$ yr, or generated during the pre-main sequence evolutionary phase through a dynamo process. Our results, and those of \cite{vaytet_2018, machida_2019, wurster_2022}\footnote{\cite{wurster_2018} previously reported that implanting a $\sim \mathrm{kG}$ field strength in the protostar was not possible, although their more recent study in \cite{wurster_2022} showed that this was due to inadequate resolution.}, show that the measured values may be achieved and maintained following the second collapse. The uncertainty here lies in the short horizon of predictability of second collapse calculations, as they must be able to simulate $\sim 10^{5}$ years following protostellar birth in order to make an adequate comparison with observations. As such, in order to determine the origin of the magnetic fields measured in YSOs, better constraints on magnetic field strengths both at dense core scales through the measurement of linearly polarized dust emissions, or at much smaller scales through Zeeman broadening, are required in Class 0 sources. However, these measurements are immensely difficult to undertake owing to the optical depths involved.
\\
In the meantime, a significant amount of theoretical modeling is required in order to describe the evolution of the protostellar magnetic field in conjunction with pre-stellar evolution models, however simplified such models are. Numerically costly simulations such as those presented in this study are immensely helpful in obtaining the initial properties and structure of the protostar, however their short horizon of predictability precludes them from definitively solving the magnetic flux problem.

\subsection{The missing mass problem}

Current observational surveys of Class 0/I disks estimate their masses to be $\sim 10^{-3}-10^{-2}\ \mathrm{M_{\odot}}$ (e.g., \citealp{tobin_2020}), which appears to be an order of magnitude lower than those predicted by theoretical studies ($\sim 10^{-2}-10^{-1}\ \mathrm{M_{\odot}}$, e.g., \citealp{machida_2011b, tsukamoto_2015, tsukamoto_2015b, tomida_2015, masson_2016, lee_2021}, see the discussions in \citealp{tsukamoto_2023}). Notwithstanding the uncertainties involved in current observational methods \citep{duy_2024}, this discrepancy has been dubbed the "missing mass problem". It has also been shown that current subgrid models aiming to emulate the sub-AU regions by replacing them with a sink particle show a strong sensitivity to the parameters chosen in said model. In \cite{hennebelle_disks}, the sink accretion threshold was shown to particularly affect the disk mass, with lower accretion thresholds leading to lower disk masses. The results of the hydrodynamical runs of \cite{ahmad_2024} seemed to show that the sink accretion threshold used in most simulations ($\approx 1.66\times 10^{-11}\ \mathrm{g\ cm^{-3}}$) is a factor $\approx 40$ lower than it should be, thus exacerbating the missing mass problem as that would mean that disks are in reality much more massive in simulations. In the present study, the inclusion of magnetic fields results in a lower disk density, and by extension a lower density measured at the disk shock front, placing it at $\approx 2\times 10^{-10}\ \mathrm{g\ cm^{-3}}$ when it reaches a radius of 1 AU, as opposed to the $\approx 5.93\times 10^{-10}\ \mathrm{g\ cm^{-3}}$ obtained in the hydrodynamical run. This result highlights the sensitivity of the nascent circumstellar disk to the magnetic field strength that is inherited by the sub-AU region. Thus, a more thorough understanding of the disk properties not only requires one to study the nascent protostar and disk in concert, but also requires better constraints on dust resistivities that dictate the amount of magnetic flux, and by extension the amount of angular momentum inherited by the disk.
\\
In any case, this issue highlights the need for better comparisons between observations and theoretical models, as \cite{duy_2024} has shown that current observational estimates of disk masses are inadequate and fail to predict the current sizes when compared to simulations. Advancements in this regard are of great importance to the field, as constraining the masses of circumstellar disks is crucial to determine the initial mass budget for planet formation. 

\subsection{The importance of adequate dust resistivity tables}

An important uncertainty in our current understanding of sub-AU regions is the dust resistivity used. The MRN dust size distribution is increasingly called into question by studies that account for dust coagulation and fragmentation during protostellar collapses \citep{lebreuilly_2023c, kawasaki_2023, tsukamoto_2023, bhandare_2024, goy_2024}. Our simulations are undertaken under the assumption that Ohmic resistivity is, as predicted by dust-size distribution studies, negligible. This leads to stronger magnetic fields within the first Larson core, which in turn increases the magnetic field intensity in the nascent protostar and circumstellar disk. As a result, magnetic torques drive considerably more material towards the protostar, thus leading to a reduced disk density.
\\
As such, the properties of the newly-formed circumstellar disk are highly sensitive to the dust resistivities that dictate the magnetic field intensity inherited from larger spatial scales. A better understanding of the sub-AU regions is predicated upon accurate dust resistivity tables, which requires a better understanding of the dust size distribution. The Hall effect however, remains an important caveat as there appears to be conflicting models in the literature regarding its effects during the collapse \citep{wurster_2021b, tsukamoto_2022, Hopkins_2024}. Progress will ultimately be achieved by longer wavelength observations of star-forming regions in order to probe optically thin dust emissions, as well as by advances in our theoretical modeling of dust growth and fragmentation during protostellar collapses. Recent numerical advances in the description of these processes may allow for their inclusion in fully 3D hydrodynamical simulations \citep{lombart_2024}.

\subsection{A magneto-rotational instability?}

\begin{figure}[h]
\centering
\includegraphics[scale=.45]{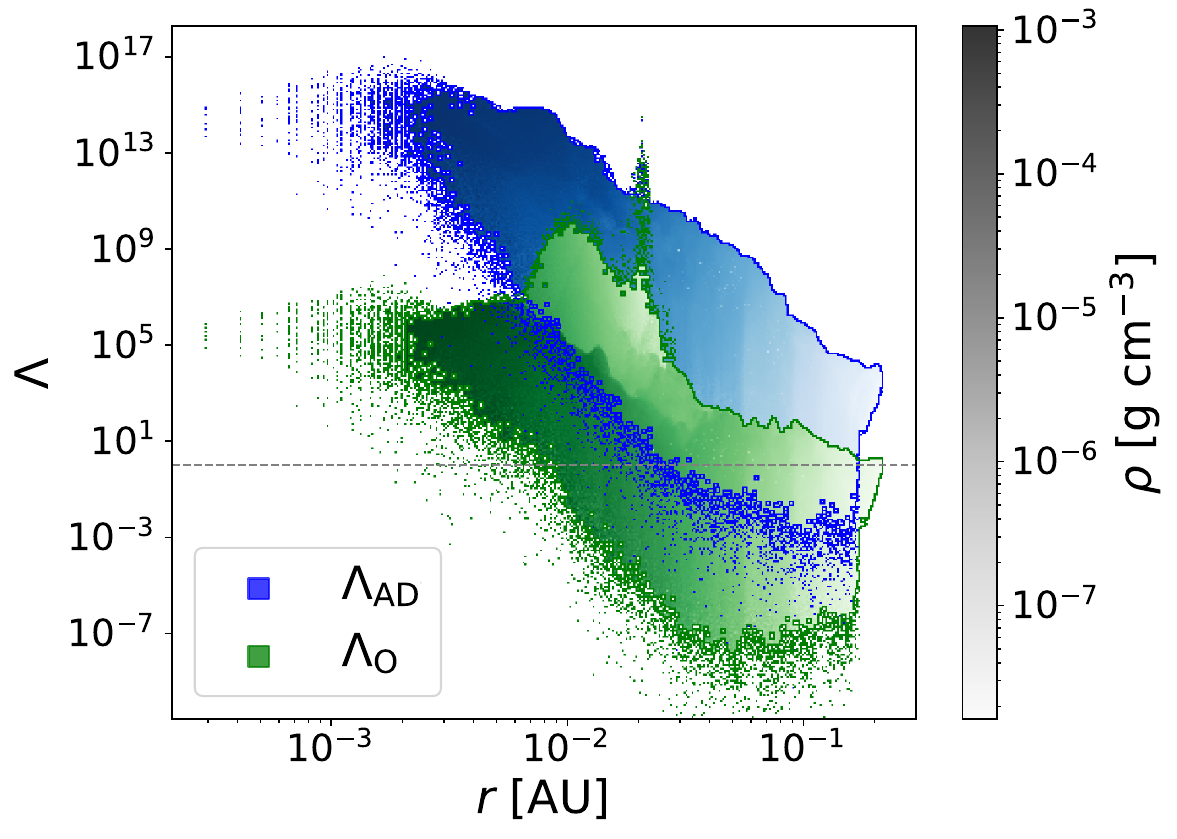} 
\caption{A set of 2D histograms binning the cells belonging to the protostar and circumstellar disk (see \hyperref[appendix:defProtostar]{Appendix \ref*{appendix:defProtostar}} for an overview of each object's definition) at the final simulation snapshot of run NIMHD ($t\approx 0.55$ yr after protostellar birth). These display the ambipolar ($\Lambda_{\mathrm{AD}}$, blue distribution) and Ohmic ($\Lambda_{\mathrm{O}}$, green distribution) Elsässer numbers as a function of radius. The gray dotted line denotes $\Lambda=1$. The color shading is related to the average density within the bin, as indicated by the gray-scale colorbar.}
\label{fig:elsasserHists}
\end{figure}

\begin{figure*}[h]
\centering
\includegraphics[scale=.4]{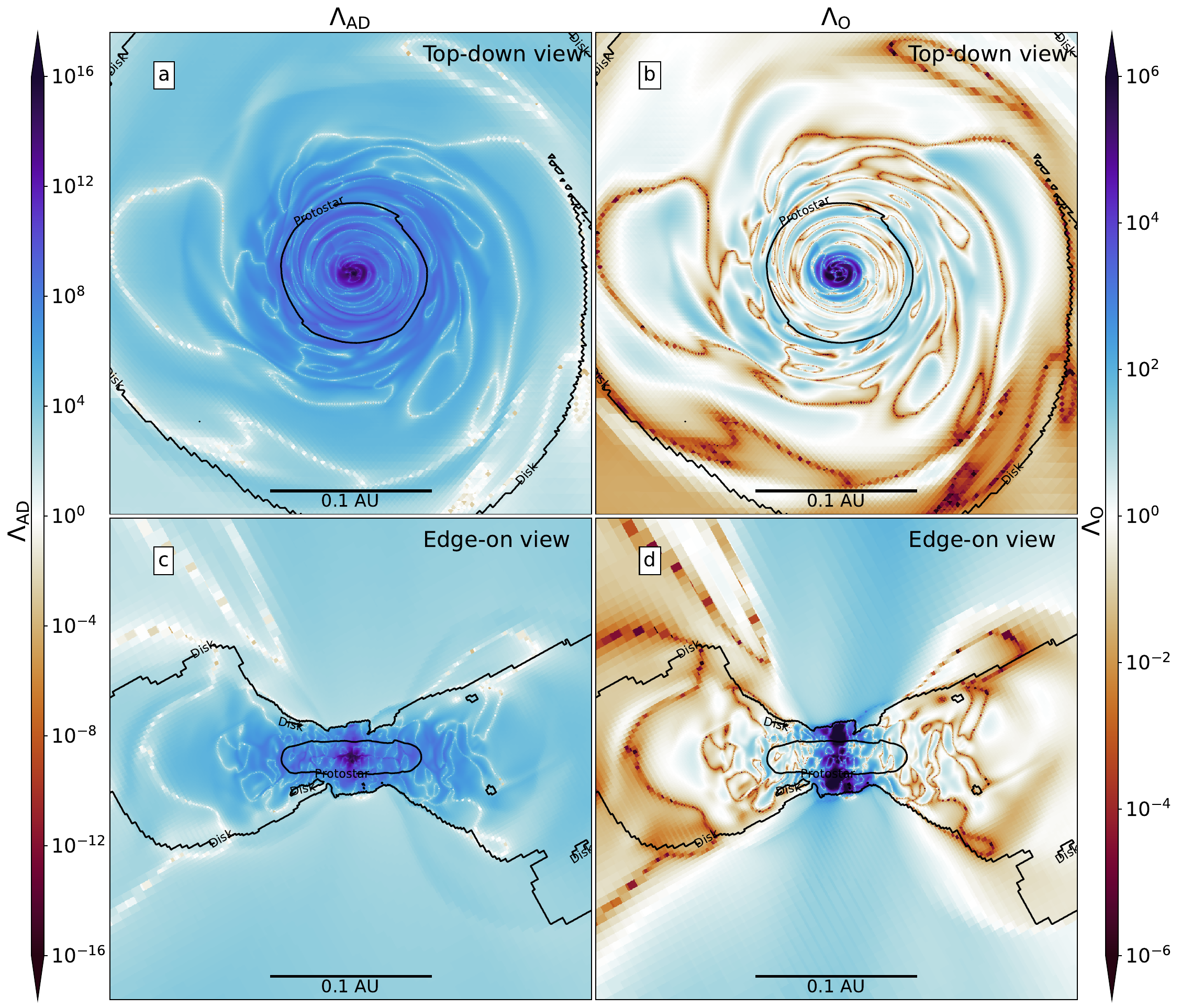} 
\caption{Slices showing the ambipolar (first column, panels a and c) and Ohmic (second column, panels b and d) Elsässer numbers within the star-disk system in the final snapshot of run NIMHD ($t\approx 0.55$ yr after protostellar birth). The first row (panels a and b) displays slices in a top-down view, whereas the second row (panels c and d) displays edge-on views. The colorbar on the left applies to panels (a) and (c), whereas the one on the right applies to panels (b) and (d). The outer contour corresponds to the disk surface, whereas the inner one corresponds to the protostar's surface (see \hyperref[appendix:defProtostar]{Appendix \ref*{appendix:defProtostar}} for an overview of each object's definition).}
\label{fig:elsasserSlices}
\end{figure*}

\begin{figure}[h]
\centering
\includegraphics[scale=.4]{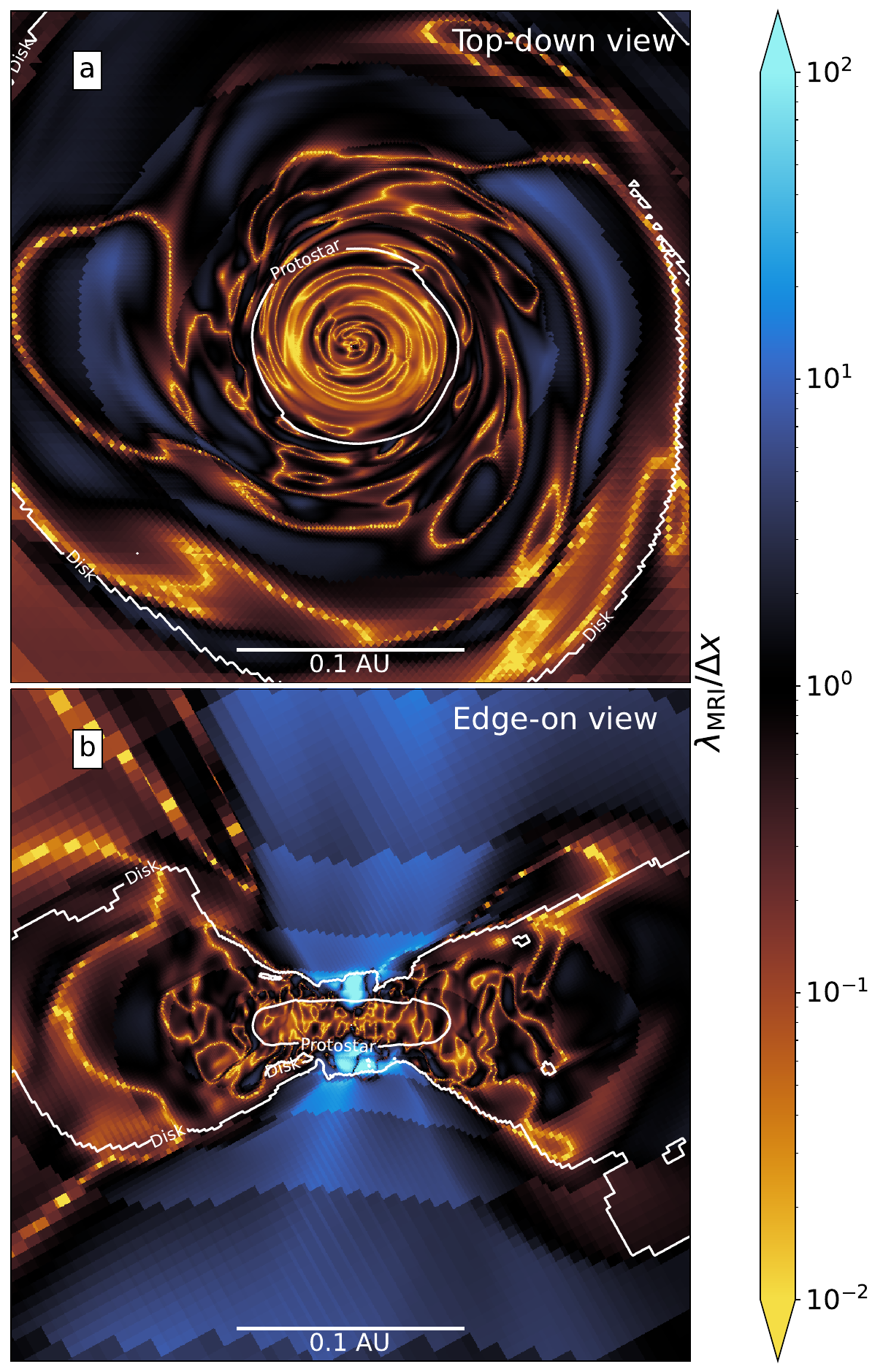} 
\caption{Top-Down (panel a) and edge-on (panel b) slices accross the star-disk system showing the ratio of the wavelength of the most unstable MRI mode to cell size at the final snapshot of run NIMHD ($t\approx 0.55$ yr after protostellar birth). The outer contour corresponds to the disk surface, whereas the inner one corresponds to the protostar's surface.}
\label{fig:lmridx}
\end{figure}

In run NIMHD, we witness a hot and highly magnetized disk ($T \sim 10^{3}$ K and $B \sim [10^2 - 10^3]$ G), see Figs. \ref{fig:rhoT}, \ref{fig:nimhdprofiles} and \ref{fig:nimhdbfield}), whose plasma $\beta$, as shown in \hyperref[fig:nimhdbfield]{Fig. \ref*{fig:nimhdbfield}} is well above unity. This naturally calls for an analysis to determine whether such a disk is prone to develop the Magneto-Rotational Instability (MRI, \citealp{balbus_1991, wardle_1999, lesur_2014, riols_2019, deng_2020}), an instability known to occur in such conditions. Although we chose to omit Ohmic dissipation from our simulation, the analysis hereafter presented also accounts for Ohmic resistivity, as it is  the non-ideal MHD effect commonly responsible for quenching MRI in other studies in the literature \citep{lesur_2014}. Hence, its inclusion in this section's analysis is purely to determine wether or not it may quench the MRI \textit{were} we to include it.
\\
In order to ascertain whether an MRI can operate in run NIMHD, we first evaluate wether the ideal MHD limit is recovered within the star-disk system. This can be done by computing the ambipolar and Ohmic Elsässer numbers (respectively $\Lambda_{\mathrm{AD}}$ and $\Lambda_{\mathrm{O}}$):
\begin{equation}
    \Lambda_{\mathrm{AD}} = \frac{v_{\mathrm{A,z}}^{2}}{\eta_{\mathrm{AD}}\Omega},
\end{equation}
\begin{equation}
    \Lambda_{\mathrm{O}} = \frac{v_{\mathrm{A,z}}^{2}}{\eta_{\mathrm{O}}\Omega},
\end{equation}
where $v_{\mathrm{A,z}}$ is the vertical Alfvén velocity, $\eta_{\mathrm{AD}}$ (respectively $\eta_{\mathrm{O}}$) is the ambipolar (respectively Ohmic) resistivity,\footnote{The ambipolar and Ohmic resistivities are obtained through an interpolation of the \citep{marchand_2016} table.} and $\Omega$ is the angular velocity. When the Elsässer number is maintained below unity, the MRI is effectively quenched \citep{jin_1996, turner_2007, simon_2018}.
\\
The resulting measurements of $\Lambda_{\mathrm{AD}}$ and $\Lambda_{\mathrm{O}}$ are presented in \hyperref[fig:elsasserHists]{Fig. \ref*{fig:elsasserHists}}, which displays the two quantities as a function of radius within the protostar and circumstellar disk in 2D histograms. The intensity of the color shade is linked to the average density within each bin. As a complement to this figure, we show in \hyperref[fig:elsasserSlices]{Fig. \ref*{fig:elsasserSlices}} both top-down (frist row) and edge-on (second row) slices displaying $\Lambda_{\mathrm{AD}}$ (first column) and $\Lambda_{\mathrm{O}}$ (second column).
\\
We see that $\Lambda_{\mathrm{AD}}$ is above unity throughout the system, reaching values of $\sim 10^{16}$. It is below unity only for low density gas at larger radii, which appear to be rather sparsely distributed, as seen in panels (a) and (c) of \hyperref[fig:elsasserSlices]{Fig. \ref*{fig:elsasserSlices}}. $\Lambda_{\mathrm{O}}$, however, is consistently above unity only in the innermost regions belonging to the protostar, and it fluctuates throughout the rest of the star-disk system according to the local magnetic field strength.\footnote{$v_{\mathrm{A,z}}\propto B_{\mathrm{z}}/\sqrt{\rho}$.} These measurements show that the ideal MHD limit is mostly recovered within the star-disk system, and should we have included Ohmic dissipation in our calculations, it would have been mostly recovered in the innermost regions of the system.
\\
In order to assess wether the conditions to trigger the MRI are met, one final criterion is necessary; a comparison of the disk scale height $h=\frac{c_{\mathrm{s}}}{\Omega}$ (where $c_{\mathrm{s}}$ is the sound speed) to the wavelength of the most unstable MRI mode ($\lambda_{\mathrm{MRI}}$, \citealp{balbus_1991}):
\begin{equation}
    \lambda_{\mathrm{MRI}} = \frac{2\pi}{\sqrt{3}}\frac{v_{\mathrm{A,z}}}{\Omega}.
\end{equation}
The MRI operates when $h>\lambda_{\mathrm{MRI}}$ \citep{balbus_1991}, a condition that is met within the disk as $h\in[10^{-3}, 10^{-1}]$ AU, whereas $\lambda_{\mathrm{MRI}}\in[10^{-10}, 10^{-2}]$ AU. As such, we conclude that the circumstellar disk born out of the second gravitational collapse is prone to triggering a dynamo process through the MRI mechanism. 
\\
However, given the small values of $\lambda_{\mathrm{MRI}}$ witnessed within the star-disk system, we do not have enough resolution to adequately resolve the instability. This is shown in \hyperref[fig:lmridx]{Fig. \ref*{fig:lmridx}}, which displays $\lambda_{\mathrm{MRI}}/\Delta x$ in slices at the final snapshot of run NIMHD. $\lambda_{\mathrm{MRI}}/\Delta x$ is consistently below unity throughout most of the circumstellar disk, and it fails to reach 10 in the midplane, which is insufficient to adequatly capture the instability \citep{Noble_2010}. The only regions where $\lambda_{\mathrm{MRI}}/\Delta x \gg 1$ are the polar regions directly above the protostar, where the angular velocity is smallest. This could be behind the previously discussed oscillations in magnetic field strength in the innermost regions.
\\
\\
In the present study, we have shown strong torque mechanisms on the circumstellar disk that are driving high mass accretion rates towards the protostar, even without the presence of the MRI. Nevertheless, the fact that the MRI may operate within this circumstellar disk is a very significant result, as it may significantly alter the structure of the magnetic field, as well as amplify its strength to equipartion values. This could later drive a high velocity jet from protostellar scales, as well as induce even stronger torques on the disk, thus causing a higher accretion rate towards the protostar. This might ultimately cause the protostar to decouple from the circumstellar disk, and truncate the latter at a magnetospheric radius, as is observed in more evolved class I systems. Capturing the MRI in a circumstellar disk formed out of the second collapse might not be viable with current computational hardware, as the characteristic scales are too difficult to resolve. However, a different numerical setup inspired by the results obtained in this study might achieve such result.

\section{Conclusion} \label{section:conclusion}

In the present study, we have undertaken radiative MHD simulations describing the collapse of a turbulent and gravitationaly unstable dense cloud core of $1\ \mathrm{M_{\odot}}$ to stellar densities, both under the ideal MHD approximation and under the non-ideal approximation in which we have accounted for ambipolar diffusion. Our stringent refinement criterion, as well as high spatial resolution, allowed us to describe the nascent protostar and circumstellar disk with unprecedented resolution. We push the calculations as far as possible in time following protostellar birth in order to study the nascent disk's expansion, reaching $\approx 0.5$ years in our high resolution run and $\approx 1.2$ years in our lower resolution run. Our results may be summarized as follows:

\begin{enumerate}[label=(\roman*)]
  \item When accounting for ambipolar diffusion, the efficacy of magnetic braking is significantly reduced toward higher density gas, which allows the nascent protostar to reach breakup velocity and shed its surface material to form a circumstellar disk around it. The nascent disk exhibits strong eccentricity (reaching values of $\sim 0.1$), and the protostar is embedded within it. The birth and early evolution of the circumstellar disk is qualitatively similar to the RHD runs presented in \cite{ahmad_2024}, as the plasma $\beta$ within the disk far exceeds unity. The nascent disk vigorously expands in the radial direction. This result carries implications for the angular momentum problem, as we show that the protostar must achieve breakup velocity in order to form a circumstellar disk. As such, angular momentum transport processes must spin-down the protostar on considerably longer timescales than the free-fall time of the dense cloud core.
  \item The magnetic field implanted in the protostar at birth has a strength of $\sim 10^{5}$~G in the ideal MHD run, which then continuously reduces to $\sim 10^{4}$ G as the simulation progresses. In the non-ideal MHD run, the implanted field has a strength of $\sim 10^{3}$ G which is maintained throughout the simulations duration. Since current observational surveys of magnetic fields in YSOs favor the fossil field hypothesis, this puts the non-ideal MHD simulation in agreement with them.
  \item The field implanted in the protostar in the non-ideal run is mostly toroidal ($B_{\phi}$), although a notable vertical component ($B_{z}$) threads the star-disk system. Within the protostar, the vertical component is significantly built-up over time.
  \item The circumstellar disk formed in our non-ideal run has a plasma $\beta$ well above unity, with a strong magnetic field whose strength ranges from $[10^{2}-10^{3}]$ G. Our analysis shows that this disk is prone to triggering a dynamo process through the magneto-rotational instability (MRI), although we do not have the resolution to adequately capture the mechanism. This would be the case even if we were to account for Ohmic dissipation. The MRI might be responsible for the decoupling of the protostar from the disk in which it is embedded, and transition the system towards a magnetospheric accretion mechanism reminiscent of class I systems.
  \item Owing to our use of turbulent initial conditions, the magnetic field mostly loses its coherence and we see no outflows or jets in both runs. In the non-ideal MHD run however, the plasma $\beta$ in the polar regions upstream of the protostellar accretion shock is continuously being reduced. Coupled with the fact that a vertical component is being built-up in the protostar, this may lead to the launching of an outflow at later times.
  \item When comparing the nascent disk in the non-ideal MHD run to its hydro counterpart (from \citealp{ahmad_2024}), we note a reduced disk density. This is caused by the presence of strong magnetic and gravitational torques within the disk, which transport a significant amount of material towards the protostar. This also causes the protostar to become more massive than in the hydro case. The reduced disk density in turn causes a reduced density at the disk's equatorial shock front, which is an important measure for studies of global disk evolution that leverage sink particles to advance the simulations in time. The trends seen in our simulations indicate that the shock front's density at 1 AU in the magnetized case is a factor $\approx 3$ lower than that reported in the hydro runs of \cite{ahmad_2024}. As such, we conclude that constraining current subgrid model parameters used in the literature require better constraints on dust resistivities, thus highlighting the need for a more comprehensive modeling of the dust size distribution throughout the collapse.
\end{enumerate}

\noindent Although we may learn a lot from expensive simulations like the ones presented in this study, it is important to note that their horizon of predictability is rather short and their results may not be applicable throughout the entirety of the Class 0 phase. The importance of magnetic fields in dictating the transport of material within the circumstellar disk also highlights the need for better constraints on the dust-size distribution, which requires significant observational and theoretical efforts.

\begin{acknowledgements}
      We thank the anonymous referee for their numerous comments that have significantly improved the quality of the manuscript. This work has received funding from the French Agence Nationale de la Recherche (ANR) through the projects COSMHIC (ANR-20-CE31- 0009), DISKBUILD (ANR-20-CE49-0006), and PROMETHEE (ANR-22-CE31-0020). We have also received funding from the European Research Council synergy grant ECOGAL (Grant : 855130). The simulations were carried out on the Alfven super-computing cluster of the Commissariat à l'Énergie Atomique et aux énergies alternatives (CEA). We thank Elliot Lynch for valuable discussions during the writing of this paper. Post-processing and data visualization was done using the open source \href{https://github.com/osyris-project/osyris}{Osyris} package.
\end{acknowledgements}

\bibliographystyle{aa}
\bibliography{biblio}

\begin{appendix}
\section{Defining the protostar and circumstellar disk} \label{appendix:defProtostar}

\begin{figure*}[h]
\centering
\includegraphics[scale=.5]{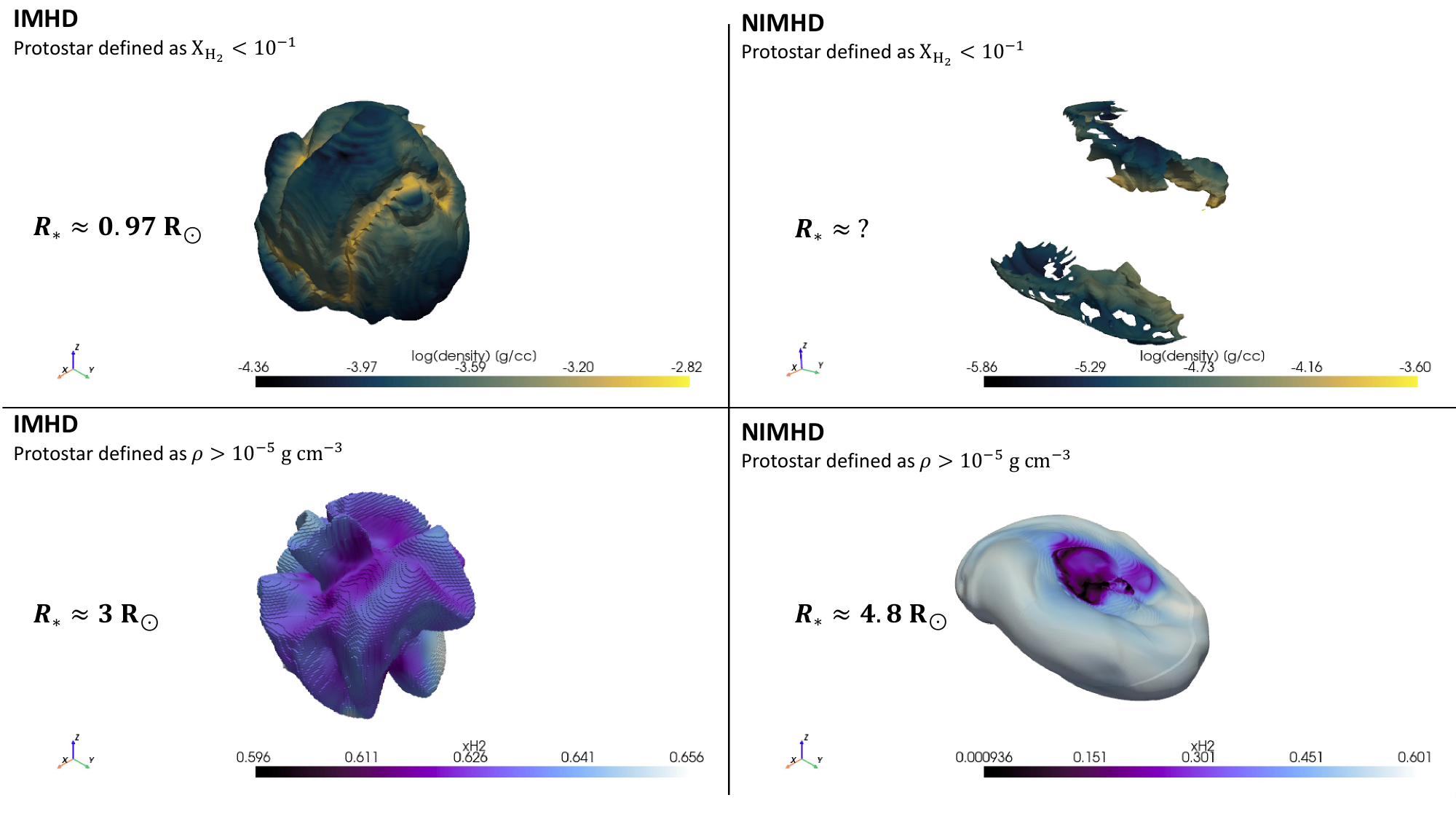} 
\caption[Illustration of stellar criterion]{A 3D illustration of the two criteria used to define the protostar, as applied in run IMHD (first columns) and run NIMHD (second columns). The first row displays an isocontour of $X_{\mathrm{H}_{2}}\approx10^{-1}$, whereas second row displays an isocontour of $\rho \approx 10^{-5}\ \mathrm{g\ cm^{-3}}$. The colorbar in the first (resp. second) row displays the gas density (resp. $X_{\mathrm{H}_{2}}$) in the extracted surface.}
\label{fig:starcriterion}
\end{figure*}

Since the two runs yield drastically different qualitative properties, one must have robust definitions for both the protostar and the circumstellar disk before proceeding to any quantitative comparison. In the case of run IMHD, the protostar is a thermally supported body, however, simply finding the cells in which thermal pressure support is attained as is done in \cite{ahmad_2023} is inadequate, as the current sheets protruding from the stellar surface also satisfy this definition. As such, we have decided to define the protostar as being all cells in which at least $90\%$ of H$_{2}$ molecules are dissociated (i.e., $X_{\mathrm{H}_2}<10^{-1}$, where $X_{\mathrm{H}_2}$ is the fraction of hydrogen under molecular form).
\\
In the case of run NIMHD, the presence of centrifugal support drastically changes the structure of the protostar, which flattens along the equator. Further complicating things, the transition from thermal pressure support to mainly centrifugal support against gravity is smooth, and no shock front separates the protostar from its circumstellar disk \citep{ahmad_2024}. As a result, we adopt the same arbitrary definition for the protostar as in \cite{vaytet_2018, ahmad_2024}, namely, that it is the gas whose density exceeds $10^{-5}\ \mathrm{g\ cm^{-3}}$. To illustrate why these two criteria were used, we display in \hyperref[fig:starcriterion]{Fig.~\ref*{fig:starcriterion}} their results when applied to both simulations. The criterion defining the protostar as being $X_{\mathrm{H}_2}<10^{-1}$ is displayed in the first row, where we see that it recovers the stellar surface in run IMHD but fails to do so in run NIMHD. On the other hand, the second criterion stating that the star is defined as $\rho>10^{-5}\ \mathrm{g\ cm^{-3}}$ and displayed in the second row, shows that it selects extended current sheets protruding from the stellar surface in run IMHD but recovers a centrifugally flattened surface in run NIMHD.
\noindent The circumstellar disk is defined as in \cite{ahmad_2024}; it is the centrifugally supported gas whose thermal pressure support exceeds incoming ram pressure, and whose density exceeds the density of the shock front (which is in turn determined through ray-tracing).

\section{Rotational breakup of the protostar} \label{appendix:breakup}

\begin{figure}[h]
\centering
\includegraphics[scale=.45]{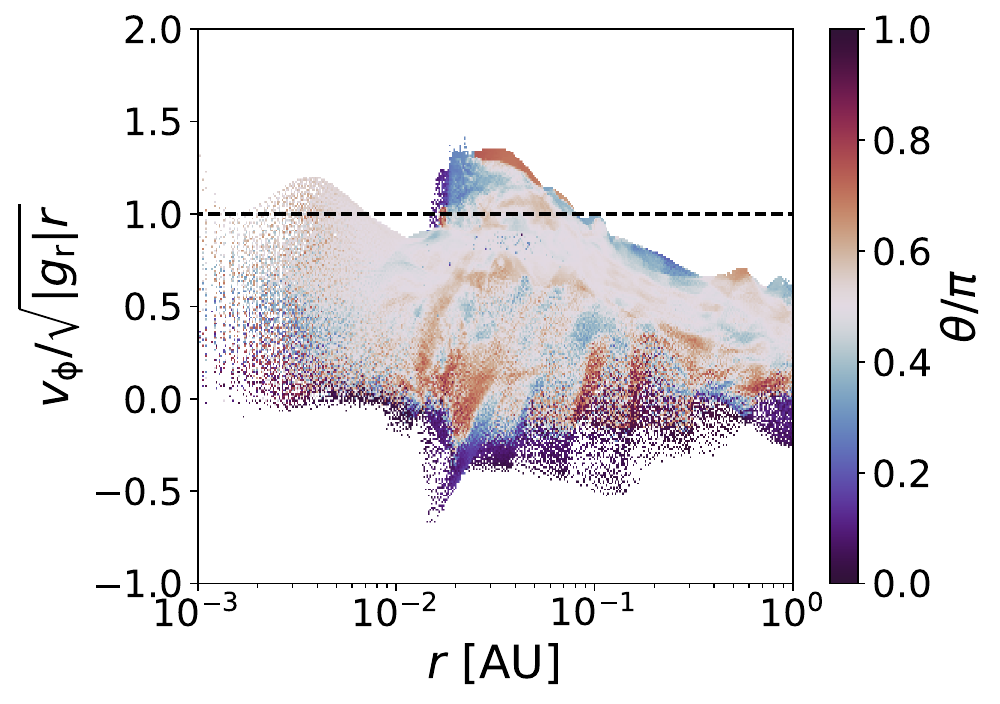} 
\caption{A 2D histogram binning all the cells in run NIMHD at $t\approx 0.3$ yr (where $t=0$ corresponds to the epoch of protostellar birth), which shows the distribution of azimuthal velocities divided by $\sqrt{|g_{\mathrm{r}}|r}$ with respect to radius. The color code in the histograms represents the co-latitude $\theta$ divided by $\pi$, where $\theta/\pi =0.5$ corresponds to the equator and $\theta/\pi = 1$ (respectively 0) corresponds to the south (respectively north) pole.}
\label{fig:breakup}
\end{figure}

Herein, we present the measurements of $v_{\mathrm{\phi}}/\sqrt{|g_{\mathrm{r}}|r}$, where $g_{\mathrm{r}}=-\partial \phi/\partial r$ ($\phi$ being the gravitational potential obtained through the Poisson equation), as is done in \cite{ahmad_2024}, in order to demonstrate that the protostar in run NIMHD undergoes a rotational breakup. The results, presented in \hyperref[fig:breakup]{Fig. \ref*{fig:breakup}}, show that parcels of fluid exceed breakup velocity, thus causing the gas to spread outwards due to excess angular momentum.

\section{An additional zoom-out to predict $\rho_{\mathrm{acc}}(1\ \mathrm{AU})$} \label{appendix:evenlowerres}

\begin{figure}[h]
\centering
\includegraphics[scale=.45]{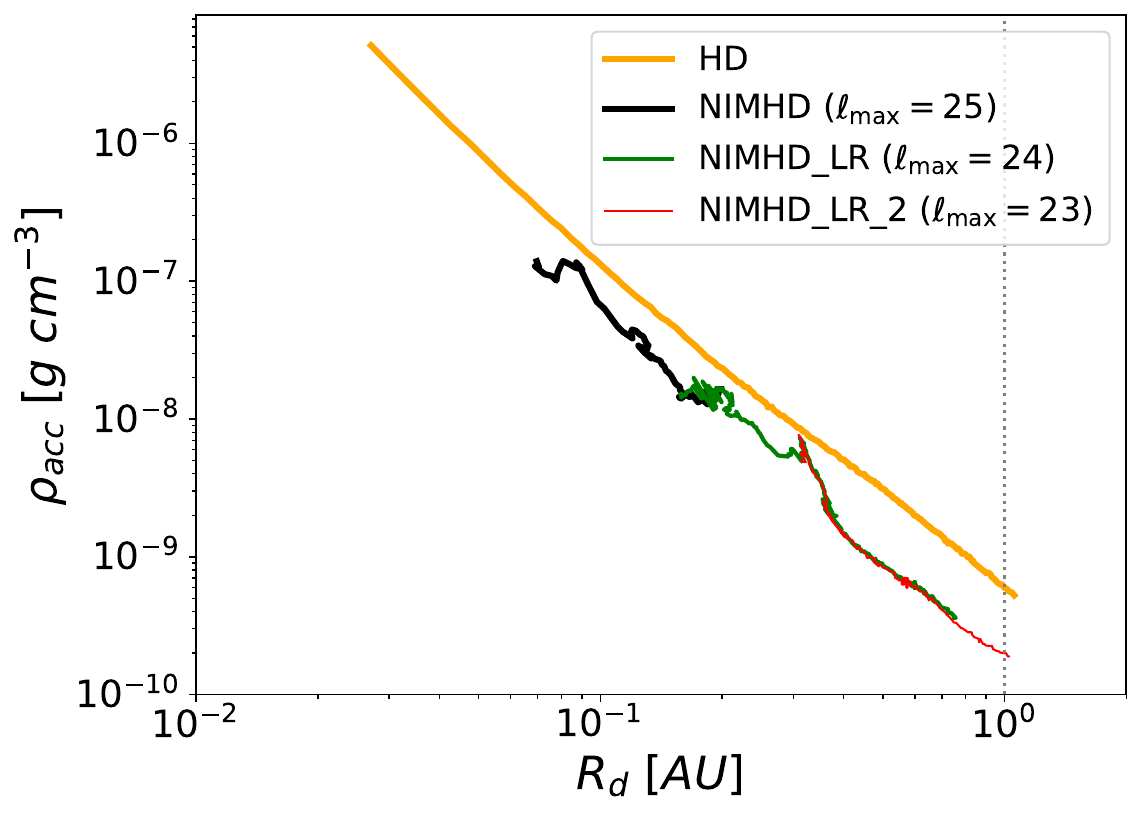} 
\caption[Illustration of stellar criterion]{Density measured at the disk's equatorial shock front as a function of the disk's equatorial radius $R_{\mathrm{d}}$, for run NIMHD (black curve) and its hydro counterpart (orange curve). The green curve is a zoom-out branched from run NIMHD and run at a lower resolution with $\ell_{\mathrm{max}}=24$ (run NIMHD\_LR), and the red curve is a zoom-out branched out of run NIMHD\_LR and run at $\ell_{\mathrm{max}}=23$ (run NIMHD\_LR\_2). The significant overlap between the results of NIMHD\_LR and NIMHD\_LR\_2 (green and red curves) shows that the results of this doubly zoomed-out run are converged with regards to $\rho_{\mathrm{acc}}$.}
\label{fig:rhoaccALL}
\end{figure}

One of the goals of this study is to provide a prediction for $\rho_{\mathrm{acc}}(1\ \mathrm{AU})$, the disk's equatorial shock front density when it reaches a size of 1 AU. Run NIMHD\_LR, although pushed much further out in time than run NIMHD, is still prohibitively expensive. As such, we have branched run NIMHD\_LR at $t\approx 0.92$ yr after protostellar birth and pushed the calculations until the disk radius reached $\approx 1$ AU. The results of this run, labeled NIMHD\_LR\_2 are displayed in the red curve of \hyperref[fig:rhoaccALL]{Fig. \ref*{fig:rhoaccALL}}.

\section{Gravitational stability of the circumstellar disk}
\label{appendix:GI}

\begin{figure*}[h]
\centering
\includegraphics[scale=.3]{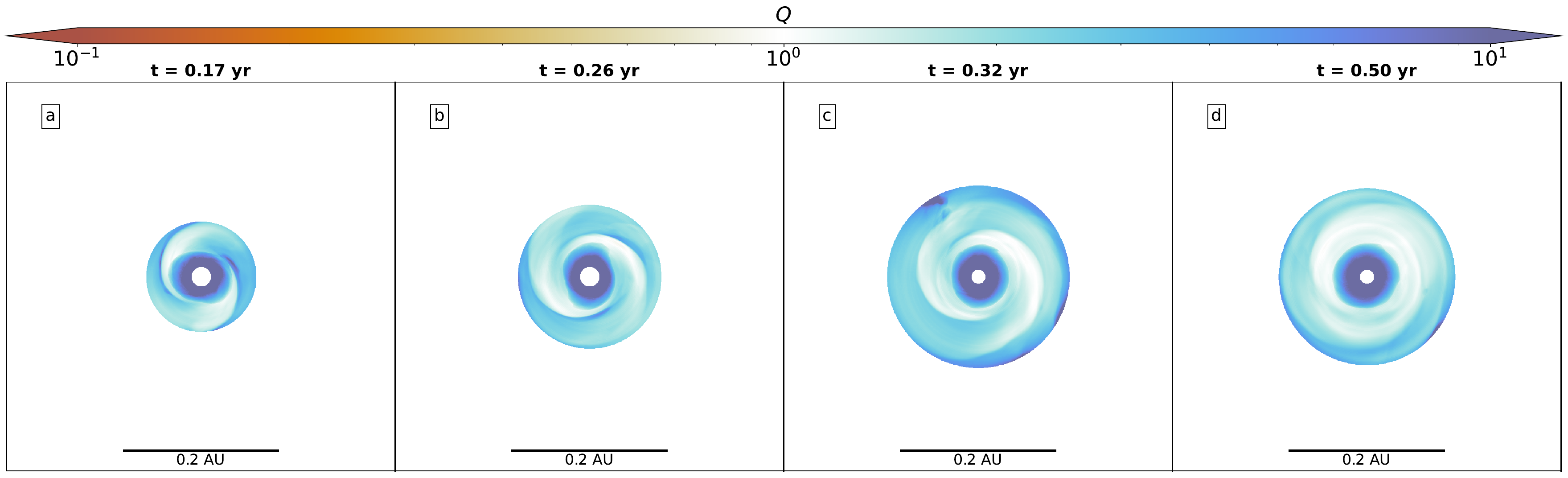} 
\caption{Real part of the Toomre $Q$ parameter of the circumstellar disk of run NIMHD, seen in a top-down view where each panel corresponds to a different time, with $t=0$ corresponding to the epoch of protostellar birth. Only cells belonging to the circumstellar disk were selected when producing these plots.}
\label{fig:toomre}
\end{figure*}

In this section, we provide measurements of the Toomre $Q$ parameter \citep{toomre_1964}
\begin{equation}
    Q = \frac{\omega c_{\mathrm{s}}}{\pi G \Sigma},
\end{equation}
where $\omega$ is the epycyclic frequency. This parameter analyses the gravitational stability of the circumstellar disk formed in run NIMHD, with $Q>1$ indicating a disk that is stable against gravitational instabilities.
\\
The resulting measurements are presented in \hyperref[fig:toomre]{Fig. \ref*{fig:toomre}}, which shows that $Q\sim 1$ throughout most of the circumstellar disk, thus indicating marginal stability, as is expected of self-gravitating accretion disks \citep{lodato_2004, lodato_2007}.

\section{Directional mass flux}
\label{appendix:massflux}

\begin{figure}[h]
\centering
\includegraphics[scale=.4]{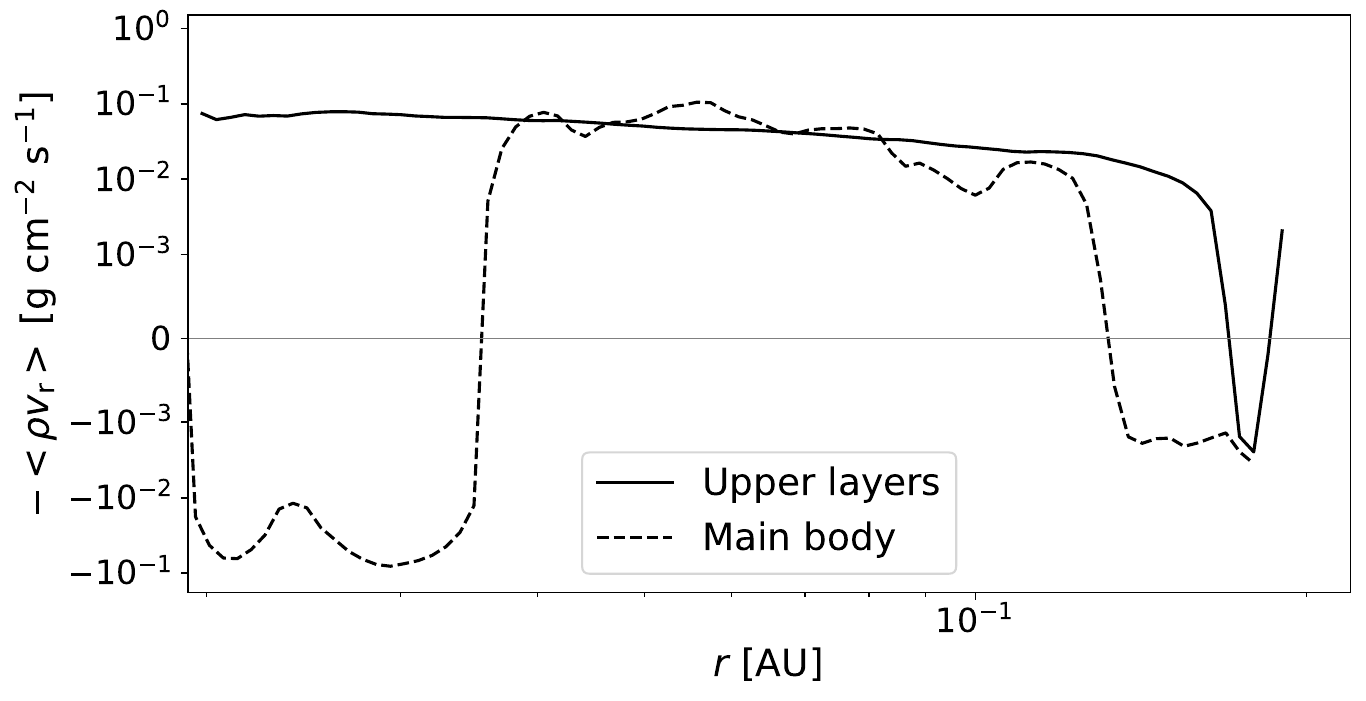} 
\caption{Average radial mass flux within the circumstellar disk in run NIMHD, measured in radial bins at $t\approx 0.5$ yr (where $t=0$ corresponds to the epoch of protostellar birth). Measurements are performed along the upper layers of the disk (solid line) and the main body (dashed line). Only cells belonging to the circumstellar disk were used in the computation.}
\label{fig:directionalmassflux}
\end{figure}

In this section, we provide measurements of the radial mass flux within the circumstellar disk in run NIMHD, both for the upper layers of the disk, and its main body. These are defined as:
\begin{equation*}
      \begin{cases}
    \text{Upper layers: Disk cells such that } \rho < 3\rho_{\mathrm{S}}, &\\
    \text{Main body: All remaining disk cells },
  \end{cases}
\end{equation*}
where $\rho_{\mathrm{S}}$ is the disk's shock density obtained through ray-tracing. This definition was chosen such that the $3\rho_{\mathrm{S}}$ contour produces a visually compelling result for the upper layers of the disk, comprising of the region just downstream of the shock front. The radial mass flux is computed using $-\rho v_{\mathrm{r}}$, which is then averaged within each radial bin. The resulting measurements are presented in \hyperref[fig:directionalmassflux]{Fig. \ref*{fig:directionalmassflux}}, which show that material is flowing inwards throughout most the disk, both in the upper layers and in the main body. The innermost regions of the main body of the disk show outward transport due to excess angular momentum.

\end{appendix}
\end{document}